\documentclass{aa}  
\usepackage{graphicx}
\usepackage{txfonts}
\usepackage{hyperref}
\hypersetup{
    colorlinks=true,
    citecolor=blue,
    linkcolor=blue,
    urlcolor=blue
    }
\usepackage{siunitx}
\usepackage{booktabs}

\newcommand*\powr{PoWR}
\newcommand*\hoii{Ho\,II \mbox{X-1}}
\newcommand*\cloudy{C{\sc loudy}}
\newcommand*\hst{HST}
\newcommand*\heiii{He\,{\sc iii}}

\newcommand*\alii{Al\,{\sc ii}\,$\lambda\lambda$\,1855,1863\,\r{A}}
\newcommand*\msun{$M_\odot$}

\begin{document}

\title{Multi-wavelength spectroscopic analysis of the ULX Holmberg II X-1 and its nebula suggests the presence of a heavy black hole accreting from a B-type donor \thanks{Based on observations made with the NASA/ESA {\em Hubble} Space Telescope, obtained from the data archive at the Space Telescope Science Institute. STScI is operated by the Association of Universities for Research in Astronomy, Inc. under NASA contract NAS 5-26555. These observations are associated with the GO programme 16182.}}

   \titlerunning{Multi-wavelength spectroscopic analysis of the ULX Holmberg II}

   %\subtitle{A comprehensive study of its emission in X-rays, UV and optical}

   \author{
        S. Reyero Serantes\inst{1}
        \fnmsep\thanks{sabela@astro.physik.uni-potsdam.de}
          \and
          L. Oskinova\inst{1}
          \and
          W.-R. Hamann\inst{1}
          \and
          V. M. G\'omez-Gonz\'alez\inst{1}
          \and
          H. Todt\inst{1}
          \and
          D. Pauli\inst{1}
          \and
          R. Soria\inst{2}
          \and
          D. R. Gies\inst{3}
          \and
          J. M. Torrej\'on\inst{4}
          \and
          T. Bulik\inst{5}
          \and
          V. Ramachandran\inst{6}
          \and
          A. A.C. Sander\inst{6}
          \and
          E. Bozzo\inst{7}
          \and
          J. Poutanen\inst{8}
%          C. Ptolemy\inst{2}\fnmsep\thanks{Just to show the usage
%          of the elements in the author field}
          }

   \institute{Institut für Physik und Astronomie, Universität Potsdam, Karl-Liebknecht-Str. 24/25, 14476 Potsdam, Germany
              %\email{wuchterl@amok.ast.univie.ac.at}
         \and
             INAF – Osservatorio Astrofisico di Torino, Strada Osservatorio 20, I-10025 Pino Torinese, Italy
        \and
             Center for High Angular Resolution Astronomy, Department of Physics and Astronomy, Georgia State University, P.O. Box 5060, Atlanta, GA 30302-5060, USA
         \and
             Instituto Universitario de F\'isca Aplicada a las Ciencias y las Tecnolog\'ias, Universidad de Alicante, 03690 Alicante, Spain
        \and
            Astronomical Observatory, University of Warsaw, Al. Ujazdowskie 4, PL-00-478 Warsaw, Poland
         \and
             Zentrum f{\"u}r Astronomie der Universit{\"a}t Heidelberg, Astronomisches Rechen-Institut, M{\"o}nchhofstr. 12-14, 69120 Heidelberg, Germany
         \and
            Department of Astronomy, University of Geneva, Chemin d’Ecogia 16, CH-1290 Versoix, Switzerland
        \and
            Department of Physics and Astronomy,  FI-20014 University of Turku, Finland
             }

   \date{Received day Month 2024; accepted day, Month year}

  \abstract 
   % context heading (optional)
   {Ultra-luminous X-ray sources (ULXs) are high-mass X-ray binaries with an X-ray luminosity above $10^{39}\,{\rm erg\,s^{-1}}$. 
   %The mechanism behind these high luminosities is still unclear.
   These ULXs can be powered by black holes that are more massive than $20M_\odot$, accreting in a standard regime, or  
   lighter compact objects accreting supercritically. There are only a few ULXs with known optical 
   or ultraviolet (UV) counterparts, and their nature is debated. Determining whether optical/UV radiation is produced by the donor star or by the accretion disc is crucial for understanding ULX physics and testing massive binary evolution. 
   }
   {
   We conduct, for the first time, a fully consistent multi-wavelength spectral analysis of a ULX and its circumstellar nebula. 
   We aim to establish the donor star type and test the presence of strong disc winds in the prototypical ULX Holmberg\,II X-1 (\hoii).   
   Furthermore, we aim to obtain a realistic spectral energy distribution of the ionising source, which is needed for robust nebula analysis. 
   }
   {
  We acquired new UV spectra of \hoii\ with the \textit{Hubble} Space Telescope (\hst ) and complemented them with archival optical and X-ray data. We explored the spectral energy distribution of the source and analysed the spectra using the stellar atmosphere code \powr\ and the photoionisation code \cloudy .
  }
   {
   Our analysis of the X-ray, UV, and optical spectra of \hoii\ and its nebula consistently explains the observations. 
   We do not find traces of disc wind signatures in the UV and the optical, rejecting previous claims of the ULX being a supercritical accretor. The optical/UV counterpart of \hoii\ is explained by a B-type 
   supergiant donor star. Thus, 
   the observations are fully compatible with \hoii\ being a close binary consisting of an $\gtrsim 66\,M_\odot$ black hole accreting matter from an $\simeq 22\,M_\odot$ B-supergiant companion. 
   Furthermore, we propose a possible evolution scenario for the system, suggesting that \hoii\ is a potential gravitational wave source progenitor.}
   {}


   \maketitle

%%%%%%%%%%%%%%%%%%%%%%%%%%%%%%%%%%%%%%%%%%%%%%%%%%%%%%%%%
%%%%%%%%%%%%%%%%%%%%%%%%%%%%%%%%%%%%%%%%%%%%%%%%%%%%%%%%%
%%%%%%%%%%%%%%%%%%%%%%%%%%%%%%%%%%%%%%%%%%%%%%%%%%%%%%%%%
%%%%% I N T R O D U C T I O N 
%%%%%%%%%%%%%%%%%%%%%%%%%%%%%%%%%%%%%%%%%%%%%%%%%%%%%%%%%
%%%%%%%%%%%%%%%%%%%%%%%%%%%%%%%%%%%%%%%%%%%%%%%%%%%%%%%%%
%%%%%%%%%%%%%%%%%%%%%%%%%%%%%%%%%%%%%%%%%%%%%%%%%%%%%%%%%

\section{Introduction}

%%%%%%%%%%%%%%%%%%%%%%%%%%%%%%%%%%%%%%%%%%%%%%%%%%%%%%%%%
%%%%% General intro
%%%%%%%%%%%%%%%%%%%%%%%%%%%%%%%%%%%%%%%%%%%%%%%%%%%%%%%%%

Ultra-luminous X-ray sources (ULXs) are binary systems 
%located outside the nuclei of their host galaxies 
with X-ray luminosities exceeding $\sim3\times10^{39}$\,$\mathrm{erg\,s^{-1}}$, which is the Eddington luminosity limit for a  20\,\msun\ black hole \citep[][]{Kaaret2017,King2023}. Typically, ULXs are situated close to 
young massive star clusters, preferably in low-metallicity galaxies \citep{Poutanen2013,Kovlakas2020}.

Initially, ULXs were considered hosts of the elusive intermediate-mass black holes. However,  the discovery of coherent pulsations in the ULX M82 X-2  proved that the compact object in this system is a neutron star \citep{Bachetti2014}. Since the apparent X-ray luminosities of ULXs are more than $100$ times larger than the Eddington luminosity of a neutron star, this led to a paradigm shift --  the study of ULXs is now largely concentrating on understanding the supercritical accretion regime. 

At the same time, the discovery of gravitational wave (GW) emission from merging black holes stirred interest in the final evolutionary stages of massive binary stars. Current binary evolution models and population synthesis codes firmly predict that black holes with masses comparable to those detected by the GW observatories ($ M_\bullet > 20$\,\msun ) should exist in binary systems, preferably in low-metallicity galaxies \citep{Fragos2013,Mondal2020}.
Some of these systems should have ULX-like X-ray luminosities when accreting close to their Eddington limit  \citep{Marchant2017, Hainich2018}.

Both models -- supercritical accretion onto light compact objects and standard accretion onto more massive black holes -- are capable of explaining the ULX luminosities. 
Therefore, to determine the true nature of compact objects in individual ULXs, a detailed multi-wavelength analysis and a comparison with the different model predictions are needed.

Hydrodynamic simulations of supercritical accretion show that large amounts of mass and angular momentum should be removed from ULXs by jets and powerful outflows from accretion discs \citep[e.g.][and references therein]{Toyouchi2024}. 
However, the simulations are not yet capable of making quantitative predictions about the outflow velocities, outward mass flux, and other key parameters. These quantities determine the multi-wavelength spectral appearance of ULXs and are required for direct comparison with observations. 
Therefore, empirical studies are needed to gauge and guide the theory. 

Observationally, the smoking gun of supercritical accretion is the spectral signature of strong outflows. 
Signatures of ultra-fast disc outflows (up to $\sim0.2$c) have already been spotted as blue-shifted absorption lines in high-resolution X-ray spectra of some ULXs \citep{Pinto2016,Kosec2018-NS}. Attempts have been made to search for such X-ray absorption features in a large sample of sources using statistical methods, but the results remain inconclusive  \citep{Kosec2018}.

Another important and promising avenue for searching the signatures of outflows is provided by UV and optical spectroscopy. The strong winds are manifested in the ultraviolet (UV) as resonance lines of metals (such as C\,{\sc iv}~$\lambda 1550$\,\r{A}) with  P\,Cygni line profiles \citep{Morton1967,Lamers1999}.
Indeed, the detection of P\,Cygni line profiles in the UV spectra of low-mass X-ray binaries (LMXBs) has been considered as proof of the presence of accretion disc winds, and consequently of supercritical accretion during X-ray outbursts \citep{CastroSegura2022}.  The signatures of disc winds were also found in optical and near-infrared (near-IR) spectra of ULXs and LMXBs, typically during outbursts, as broad emission lines of hydrogen, helium, and metals, resembling characteristics of a Wolf-Rayet (WR)-type spectrum \citep{Zhou2023,Munozdarias2020,Panizo-Espinar2022,Sanchez-Sierras2023}. 

Thus, over the last decade, an apparent consensus has emerged in the literature: the detection of wind signatures in ULX spectra proves that the compact object in the source accretes supercritically  \citep[e.g.][]{Fabrika2015}. However, it remains unclear whether the 
absence of wind signatures implies that a source accretes in a standard regime, implying that the compact object is a black hole with $M_\bullet >20M_\odot$.
Additional, circumstantial evidence is needed to address this problem. 

Such evidence is provided by considering the nature and ionisation mechanisms of the circumstellar nebulae, which are commonly found around ULXs \citep{PakullMirioni2002}. Strong outflows, characteristic of supercritical accretors, should leave their fingerprints on the ULX nebulae. In contrast, if there are no strong outflows, the nebulae should be largely photoionised by the intense X-ray radiation field of the ULX.

The situation is especially complex when the donor star in a ULX has an OB or WR spectral type. The OB and WR stars are hot,  emit most of their light in the UV, and drive powerful stellar winds.
If, as binary evolution models predict, the donor star in a ULX is a hot massive star, then its radiation must significantly contribute to or even dominate the binary UV and optical spectra, including wind lines. 
Previously, the UV spectra of ULXs have been interpreted as originating in the donor star winds rather than in the accretion disc wind \citep{Long2002, Liu2004}.  
This aligns with UV and optical observations of high-mass X-ray binaries \citep[HMXBs,][]{Ramachandran2022}. In HMXBs, the accretion proceeds at a sub-Eddington rate and their UV and optical spectra can be safely attributed to the donor. 

There are only a few ULXs with securely identified counterparts in the optical and UV. 
The question arises as to whether optical and UV counterparts of ULXs should be attributed to their donor stars or to their accretion discs, or whether perhaps they have contributions from both. 
Joint multi-wavelength analysis of donor star, accretion disc, and nebula is required to answer this fundamental question and provide a comprehensive understanding of ULXs. Such an analysis is attempted in this paper.

%--------------------------------------------------------
%---- IMAGEN NEBULOSA + HST ACQ
%--------------------------------------------------------
\begin{figure}[t!]
\centering
\includegraphics[width=\linewidth]{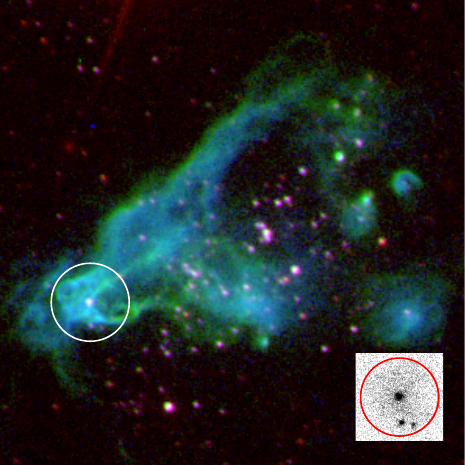}
%\caption{Legend (350 words max). Example legend text.}
\caption{
False-colour HST ACS image of the area around the ULX \hoii\ with the adjacent H\,{\sc ii} region known as the `Foot Nebula'. This composite image comprises observations taken with the F502N (blue), F658N (green), and F814W (red) filters. The white circle highlights the \heiii\ bubble generated by ionising X-ray radiation emitted by the ULX, and has a diameter of 2\farcs5, corresponding to 37\,pc at the distance of the Holmberg\,II galaxy. The ULX optical counterpart is seen as the white spot centred in the cycle. In the lower right corner, the acquisition image of our \hst\ COS observations is displayed. The 2\farcs5 diameter aperture (shown as a red circle) encompasses the ULX, the \heiii\ bubble, and two neighbouring UV-bright stars. North is up and east is to the left.
}
\label{fig:hstneb}
\end{figure}

As the object of our study, we chose Holmberg\,II X-1 (\hoii\ hereafter) -- a well-studied template ULX located at a distance of 3.05\,Mpc in a star-forming region of a dwarf galaxy with low-metallicity  \citep{Hoessel1998,Kaaret2004}. \hoii\ is surrounded by a large photoionised nebula, the Foot Nebula (Fig.\,\ref{fig:hstneb}), and is situated near a young cluster of OB-type stars from which it likely escaped \citep{Stewart2000, PakullMirioni2002,Kaaret2004, Egorov2017}. It is a perfect ULX for a multi-wavelength analysis due to its comparatively bright optical counterpart and low foreground extinction.

\hoii\  has an X-ray luminosity ($0.3-10\,{\rm keV}$) that varies from $>10^{39}$ to $\sim3\times10^{40}\,{\rm erg\,s^{-1}}$ \citep{Grise2010,Gurpide2021A}. Its X-ray properties are typical for a variable soft ULX \citep{Gurpide2021B}. \hoii\ is not an X-ray pulsar and no definite evidence of fast outflows has been found in its  X-ray spectra, which rules out that \hoii\ is accreting at a high super-Eddington rate \citep{Kosec2018,Barra2023}. 

Its optical counterpart and the circumstellar nebula were first reported by \citet{PakullMirioni2002}, and it was suggested that the nebula is photoionised by the ULX. 
Subsequent analysis set the bolometric luminosity of the source at $\sim10^{40}\,{\rm erg\,s^{-1}}$, and suggested an OB-type donor star or an X-ray illuminated accretion disc as the origin of the optical counterpart \citep{Kaaret2004,Berghea2010I,Berghea2010II}. 
Furthermore, \citet{Kaaret2004} reinforced the hypothesis of photoionisation by the examination of the relative morphologies of the He\,{\sc ii}\,$\lambda$\,4686\,\r{A}, H\,$\beta$, and [O\,{\sc i}]\,$\lambda$\,6300\,\r{A} emission lines. 
These nebular analyses, however, have never been done consistently with the analyses of the ionising source.

\citet{Fabrika2015} presented an optical spectrum of \hoii\ obtained with the Faint Object Camera And Spectrograph (FOCAS) at the 8.2\,m telescope \textit{Subaru}.
The authors claimed that the spectrum has a sufficiently high signal-to-noise ratio (S/N) for spectral analysis and fully attributed it to the accretion disc. They also claimed the detection of disc wind signatures in the spectra, which was considered as proof that \hoii\ and other ULXs accrete supercritically (see Sect.\,\ref{sec:uvlines} for a more detailed discussion of their results). In this paper, we retrieve and analyse the same spectra, complementing them with newer data. 

At radio wavelengths,  \citet{Cseh2014,Cseh2015} discovered a collimated jet emission from \hoii , which first had a kinematic luminosity of $\sim10^{39}\,{\rm erg\,s^{-1}}$ but faded out by a factor of $\sim7$ during the $\sim1.5$ years between the observations. 
\citet{Heida2014} reported the detection of the near-IR counterpart of \hoii, which \citet{Lau2017} attributed to the circumstellar disc of a B[e]-type donor star. 

In the UV, \hoii\ has been observed twice. 
\citet{Tao2012} presented \textit{Hubble} Space Telescope (\hst ) observations conducted with the prism grating of the Solar Blind Channel (SBC) of the Advanced Camera for Surveys (ACS). 
They acquired the spectrum of the ULX and two neighbouring stars (see Fig.\,\ref{fig:hstneb}). 
However, the resolving power of this instrument is not sufficient to resolve spectral lines, preventing a detailed spectroscopic analysis.

\citet{Vinokurov2022} used the UV AstroSat telescope to obtain UV photometry of \hoii . Due to the low spatial resolution of the instrument, the data are likely highly contaminated by the surrounding nebula and the contribution from the neighbouring stellar cluster.  To mitigate this, the authors performed a dedicated background subtraction; however, their reported values do not agree with the fluxes measured with the \hst\ \citep{Tao2012}. 

In both works, UV and optical photometry were combined to construct the spectral energy distribution (SED) and compared to theoretical models of irradiated accretion discs and various donor stars to elucidate the nature of the UV and optical counterpart of \hoii . 
Both studies remain inconclusive, claiming that the SED is consistent with a disc or stellar origin.

In this paper, we aim to consistently analyse the optical and UV observations, modelling the source's emission using stellar atmosphere and photoionisation models.
In Sect.\,\ref{sec:obs+datared}, we report all the data used for the analysis, as well as the data reduction process. In Sect.\,\ref{sec:meth+ana}, a detailed overview of the modelling process is given, and in Sect.\,\ref{sec:results} we present our results. Finally, we discuss the results and present an evolutionary scenario in Sect.\,\ref{sec:disc}. Finally, in Sect.\,\ref{sec:conc} we summarise our conclusions.

%%%%%%%%%%%%%%%%%%%%%%%%%%%%%%%%%%%%%%%%%%%%%%%%%%%%%%%%%
%%%%%%%%%%%%%%%%%%%%%%%%%%%%%%%%%%%%%%%%%%%%%%%%%%%%%%%%%
%%%%%%%%%%%%%%%%%%%%%%%%%%%%%%%%%%%%%%%%%%%%%%%%%%%%%%%%%
%%%%% O B S E R V A T I O N S
%%%%%%%%%%%%%%%%%%%%%%%%%%%%%%%%%%%%%%%%%%%%%%%%%%%%%%%%%
%%%%%%%%%%%%%%%%%%%%%%%%%%%%%%%%%%%%%%%%%%%%%%%%%%%%%%%%%
%%%%%%%%%%%%%%%%%%%%%%%%%%%%%%%%%%%%%%%%%%%%%%%%%%%%%%%%%

%--------------------------------------------------------
%---- IMAGEN DISENTANGLING
%--------------------------------------------------------
\begin{figure*}[t!]
\centering
\includegraphics[width=\linewidth]{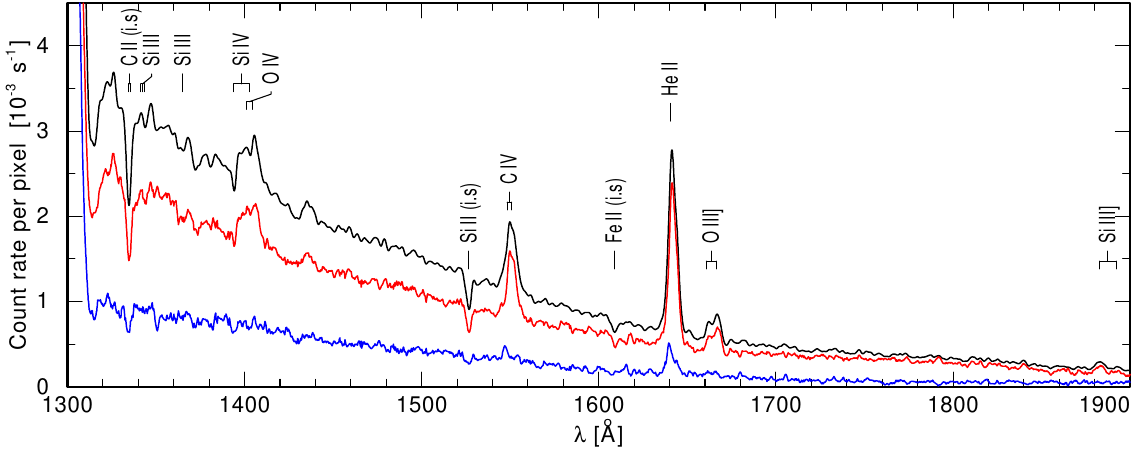}
\caption{Ultraviolet spectra extracted from the \hst\ COS observations of \hoii . 
As is shown in the insert in Fig.\,\ref{fig:hstneb}, there are three point-like sources within the COS aperture. The black line shows the combined spectrum of all three sources. 
The spectra are disentangled (Sect.~\ref{sec:uvreduce}) to separate the ULX from the neighbouring sources, presumably OB-type stars. The red line shows the spectrum of the UV counterpart of \hoii , while the combined spectra of two other sources within the COS field of view are plotted in blue.
}
\label{fig:disentangling}
\end{figure*}

\section{Observations and data reduction}
\label{sec:obs+datared}

%%%%%%%%%%%%%%%%%%%%%%%%%%%%%%%%%%%%%%%%%%%%%%%%%%%%%%%%%
%%%%% OBSERVACIONES UV 
%%%%%%%%%%%%%%%%%%%%%%%%%%%%%%%%%%%%%%%%%%%%%%%%%%%%%%%%%
\subsection{Ultraviolet spectroscopy}
\label{sec:uvreduce}

The spectroscopic UV observations of \hoii\ were carried out during February and March 2021, in the {\sc time-tag} mode with the \textit{Cosmic Origin Spectrograph} \citep[COS,][]{2012ApJ...744...60G}, onboard the \hst\ (Programme ID: 16182, PI: L. Oskinova). The low-resolution (R$\sim$\,2000) G140L grating centred at 1280\,\r{A} was used to ensure a good wavelength coverage (1230--2050\,\r{A}). The spectrum was secured over 17 orbits, grouped in five visits (Table\,\ref{table:fitsheader_COS}).

Next to the ULX, still within the COS aperture (white and red circles in Fig.\,\ref{fig:hstneb}), there is a pair of point-like sources that are only a bit fainter in the UV than our target (in the following, we consider our target as source\,1 and the narrow pair that is about 0$\farcs$8 to the south as source\,2). The focus of \hst 's main mirror suffers from spherical aberration, which is not corrected for the beam that feeds the COS spectrograph. Although we already restricted the scheduling of our observations such that the roll-angle of the spacecraft yielded the maximum separation between the two sources in the cross-dispersion direction, the spectra partially overlap on the detector. As the data reduction pipeline does not separate the sources, we developed a dedicated algorithm to disentangle their spectra. 

We employed the COS pipeline to obtain the flat-fielded detector images. These 2-D images have the wavelength dispersion as the $x$-axis and the position in the cross-dispersion direction as the $y$-axis. In these images, the positions of two maxima can be identified in the cross-dispersion direction, separated by seven pixels according to the separation of the two sources, but largely overlapping. 

We downloaded the calibration file from the \hst\ database for a single-point source with the same spectrograph set-up to obtain the cross-dispersion instrumental profile. 
This allowed us to construct a mock instrumental profile for our pair of sources with their given position and separation. We kept this profile normalised, so the only free parameter during the process is the count ratio between the two sources, $q(x)$. At each $x$-coordinate, the brightness ratio varies, searching for the value at which the deviation from the observation ($\chi^2$) is smallest. $q(x)$ is applied to the pipeline-reduced total spectrum, yielding the spectra of both sources separately (Fig.\,\ref{fig:disentangling}).

However, the faintness of the ULX in the UV limits our method. Although we have already added all 17 exposures (weighted by their respective exposure time, Table\,\ref{table:fitsheader_COS}), the number of counts per detector pixel is still very small. Therefore, the brightness-ratio function $q(x)$ has its own uncertainty which impacts the disentangled spectra. To reduce noise, we binned the images (both observation and calibration ones) in the wavelength direction by convolution with a Gaussian, and thus degrading the spectral resolution. 
For the blue part (1300 -- 1900\,\r{A})  of our spectrum, an additional smoothing with 2\AA\ is a good compromise; at longer wavelengths, the instrumental profile degrades \citep{2022cosi.book...14J} and the signal is anyhow too low to result in useful data, disentangled or not.

%%%%%%%%%%%%%%%%%%%%%%%%%%%%%%%%%%%%%%%%%%%%%%%%%%%%%%%%%
%%%%% OBSERVACIONES UV (fotometria)
%%%%%%%%%%%%%%%%%%%%%%%%%%%%%%%%%%%%%%%%%%%%%%%%%%%%%%%%%
\subsection{Ultraviolet photometry}
We obtained UV photometric measurements of \hoii\ using archival \hst\ observations taken with the prism grating PR130L ($\lambda_{\rm eff}=1761$\,\r{A}, $\Delta\lambda=[1271-1981]$\,\r{A}) of the SBC/ACS on November 27 2006 \citep[Proposal ID: 10814, PI: J. Bregman, previously analysed in][]{Tao2011}.
We used the drizzled images obtained following the standard pipeline calibration. The photometry was performed using the \texttt{APPHOT} package in \texttt{IRAF} with an aperture of 0\farcs25. For the background subtraction, an annulus around the source of 0\farcs3 was taken.
The UV counterpart of \hoii\ has a UV magnitude of $18.70\,{\rm mag}$, while each of the other two sources within the \hst\ COS aperture is roughly one magnitude fainter (Fig.\,\ref{fig:hstneb}).
In order to compare with the UV observations, in which the nebula cannot be subtracted, we also measured the total UV flux of the point-like source, including the nebula contribution (see Table\,\ref{table:photflux}).

%%%%%%%%%%%%%%%%%%%%%%%%%%%%%%%%%%%%%%%%%%%%%%%%%%%%%%%%%
%%%%% OBSERVACIONES OPTICO (espectro)
%%%%%%%%%%%%%%%%%%%%%%%%%%%%%%%%%%%%%%%%%%%%%%%%%%%%%%%%%
\subsection{Optical spectroscopy}
\label{sec:optical}

To complement the UV observations, optical archival data were retrieved. Ho\,II X-1 has been observed only three times with long-enough exposure times to ensure a sufficiently high S/N and resolution to perform a spectroscopic analysis. 

FOCAS observed the ULX, between February 26 and March 1 2011 (Proposal ID: o11104, PI: Yoshihiro Ueda). The telescope was operated with the 300B grism without a filter and a 0\farcs4 slit due to the good seeing. With this configuration, a resolving power of $R\sim$1000 was achieved for the spectral range of 3800 - 7000\,\r{A}, and an angular resolution of 0\farcs1 per pixel.  

The two other observations were performed by the \textit{Gran Telescopio de Canarias} (GTC), between November and January during the winter campaigns of 2010-2011 and 2011-2012, with the OSIRIS long-slit spectrograph (Programme IDs: GTC20-11B and GTC38-10B, PI: F. Vilardell). The instrument was operated with a 0\farcs6 slit and the R2000B grism, which has a spectral range of 3950-5700\,\r{A}.
With this configuration, a spectral resolving power of $R\sim$2165 and an angular sampling of 0\farcs25 per pixel were achieved. 

All the obtained spectra were reduced using standard {\sc iraf} tasks; for the GTC data, the \texttt{\sc gtcmos} pipeline \citep{2016MNRAS.460.1555G} was employed. 
During the process, we also corrected the data for Galactic extinction.
Due to the strong nebular contribution in the ULX spectra, the background subtraction was carried out in three independent ways, following three different approaches. These were applied to all the optical data. 

In the first one, the spectrum of the ULX is extracted only from the point-like source at the coordinates of \hoii . 
On the detector image, a narrow region contains the continuum of the point-like source, while the emission lines are extended. 
To subtract the contribution from the innermost nebula regions close to the ULX, on the detector image we select narrow regions adjacent to the continuum but still covering the emission lines. The rationale for this approach is that the ionisation conditions within the nebula strongly depend on the distance from the ionising source.  After the background subtraction, no emission nor absorption lines are present in the resulting spectra, neither for the \textit{Subaru} nor for the GTC observations.

For the second approach, we checked whether the emission line of He\,{\sc ii}\,$\lambda$\,4686\,\r{A} has a composite profile formed by the superposition of a broad wind line and a narrow nebular line. To achieve this, the spectra are extracted from a region comprising the point-like source plus the He\,{\sc ii}\,$\lambda$\,4686\,\r{A} emission. We noted that on the detector image, the extent of the He\,{\sc ii}\,$\lambda$\,4686\,\r{A} line is smaller compared to the extent of hydrogen lines, but has similar extent as the photoionised He\,{\sc iii} region \citep{PakullMirioni2002}. 
We found, however, that the He\,{\sc ii}\,$\lambda$\,4686\,\r{A} line, similar to other nebular lines, is described by a Gaussian profile with a width corresponding to the resolution of the spectrograph. This procedure is also repeated for other nebular lines, such as the forbidden [O\,{\sc iii}]\,$\lambda$\,5007\,\r{A}, with the same result.

As a third and last approach, the spectrum from a region corresponding to the whole COS 2\farcs5 aperture is extracted, and the contribution of the adjacent nebular region is subtracted. This allows a consistent analysis with the \hst\ spectrum.

%%%%%%%%%%%%%%%%%%%%%%%%%%%%%%%%%%%%%%%%%%%%%%%%%%%%%%%%%
%%%%% OBSERVACIONES X-RAY
%%%%%%%%%%%%%%%%%%%%%%%%%%%%%%%%%%%%%%%%%%%%%%%%%%%%%%%%%
\subsection{X-rays}

To model the nebular spectra it is necessary to take the ionising X-ray spectrum of \hoii\ into account.
%Ho\,II X-1 has been extensively observed in X-rays. 
In this work, we use archival observations by the \textit{XMM-Newton} telescope taken fortuitously on the same dates as our \hst\ COS observations on March 19 2021 (ObsID:0864550501, PI: M. Middleton).
The \textit{XMM-Newton} telescope delivers pipeline-reduced data ready for analysis. Since in this work, we need only the number of ionising photons and the crude spectral shape, the pipeline data are sufficient.

%%%%%%%%%%%%%%%%%%%%%%%%%%%%%%%%%%%%%%%%%%%%%%%%%%%%%%%%%
%%%%% IMAGEN COS SPECTRUM + MODELS
%%%%%%%%%%%%%%%%%%%%%%%%%%%%%%%%%%%%%%%%%%%%%%%%%%%%%%%%%

\begin{figure*}[t!]
\centering
\includegraphics[width=\textwidth]{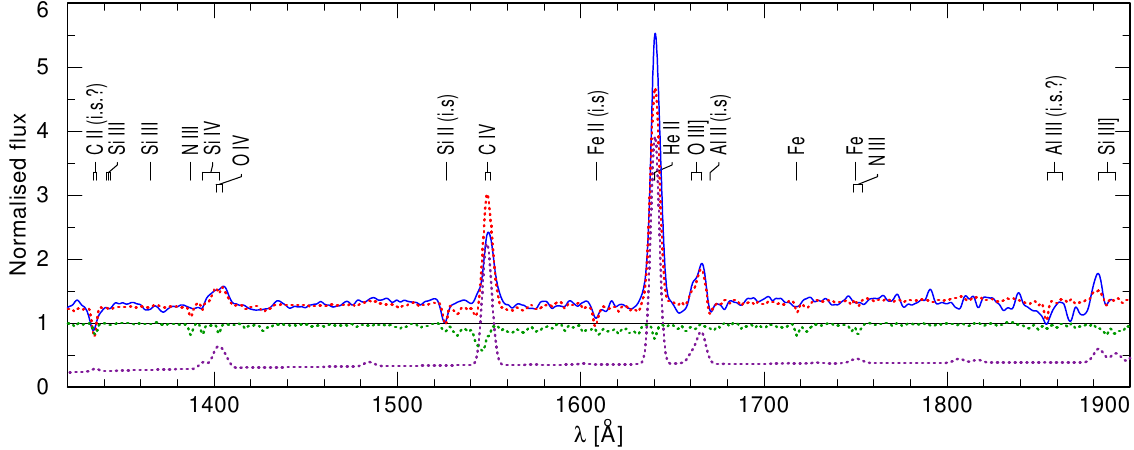}
%\caption{Legend (350 words max). Example legend text.}
\caption{Ultraviolet spectrum of \hoii\ normalised to the stellar model continuum.
Solid blue: the COS disentangled spectrum; dotted green: the modelled \powr\ stellar spectrum of a B0.5I supergiant (${T_\ast}=\,\mathrm{26\,k\,K}$, $\log g=\,3.1\,\mathrm{[cm\,s^{-2}]}$ and $\log L/L_\odot=\,5.3$, see Sect.\,\ref{sec:powr}); dotted purple: the \cloudy\ nebular model for an X-ray ionising source modelled with \texttt{diskir} (see Sect.\,\ref{sec:cloudy}); dotted red: the combined \powr\ and \cloudy\ spectra. This last one also accounts for interstellar absorption lines (i.s.)}
\label{fig:uv+models}
\end{figure*}

%%%%%%%%%%%%%%%%%%%%%%%%%%%%%%%%%%%%%%%%%%%%%%%%%%%%%%%%%
%%%%%%%%%%%%%%%%%%%%%%%%%%%%%%%%%%%%%%%%%%%%%%%%%%%%%%%%%
%%%%%%%%%%%%%%%%%%%%%%%%%%%%%%%%%%%%%%%%%%%%%%%%%%%%%%%%%
%%%%% M E T H O D S
%%%%%%%%%%%%%%%%%%%%%%%%%%%%%%%%%%%%%%%%%%%%%%%%%%%%%%%%%
%%%%%%%%%%%%%%%%%%%%%%%%%%%%%%%%%%%%%%%%%%%%%%%%%%%%%%%%%
%%%%%%%%%%%%%%%%%%%%%%%%%%%%%%%%%%%%%%%%%%%%%%%%%%%%%%%%%

%%%%%%%%%%%%%%%%%%%%%%%%%%%%%%%%%%%%%%%%%%%%%%%%%%%%%%%%%
%%%%% TABLA FOTOMETRIA
%%%%%%%%%%%%%%%%%%%%%%%%%%%%%%%%%%%%%%%%%%%%%%%%%%%%%%%%%
\begin{table}[b!]
%\footnotesize\setlength{\tabcolsep}{5pt}
\centering
\caption{
\hst\ photometric data used to construct the SED
}\label{table:photflux}
\begin{tabular}{lcc}
\hline\hline
\rule{0cm}{2.5ex}
Filter   &   Flux ULX   &  Flux ULX+nebula  \\
   &  ${[\rm 10^{-17}\,erg\,s^{-1}\,cm^{-2}]}$   &   ${[\rm 10^{-17}\,erg\,s^{-1}\,cm^{-2}]}$\\
\midrule
\rule{0cm}{2ex}F165LP  &  $12.02\pm0.02$ $^{(a)}$ &   $17.22\pm0.03$ \\
F336W   &   $2.70\pm0.54$ $^{(b)}$ &   $3.44\pm0.69$  \\
F450W   &   $1.34\pm0.27$ $^{(b)}$ &   $2.01\pm0.40$  \\
F555W   &   $0.77\pm0.16$ $^{(b)}$ &   $1.12\pm0.23$   \\
F814W   &   $0.27\pm0.05$ $^{(b)}$ &   $0.32\pm0.06$ \\
\bottomrule
\end{tabular}
\tablefoot{$^{(a)}$ Flux obtained in this work. We estimate the errors assuming Poisson distribution. $^{(b)}$ Fluxes reported in \citet{Tao2012}.
The errors in the third column are likely over-estimated, as they are obtained by propagating those in the second one. Therefore, the photometric fluxes are probably more accurate.
%$^{(a)}$ Flux for \hoii\ obtained in this work (F165LP) and by \citet{Tao2012} after performing a nebular substraction.
%$^{(b)}$ Same as column $^{(a)}$. The errors in this column are obtained by propagating the errors in column $^{(a)}$. Therefore, they only represent upper limits. 
%$^{(a)}$ \hoii\ UV flux obtained in this work. $^{(b)}$Flux for \hoii\ obtained by \citet{Tao2012} after performing a nebular substraction. The authors also report the nebular contribution on each filter. Thus, the flux values for the ULX and the nebula are recovered and presented in the second column without errors.
}
\end{table}

%%%%%%%%%%%%%%%%%%%%%%%%%%%%%%%%%%%%%%%%%%%%%%%%%%%%%%%%%
%%%%% IMAGEN SUBARU OPTICO SPECTRUM + MODELS
%%%%%%%%%%%%%%%%%%%%%%%%%%%%%%%%%%%%%%%%%%%%%%%%%%%%%%%%%
\begin{figure*}[h!]
\centering
\includegraphics[angle=90 , width=\linewidth]{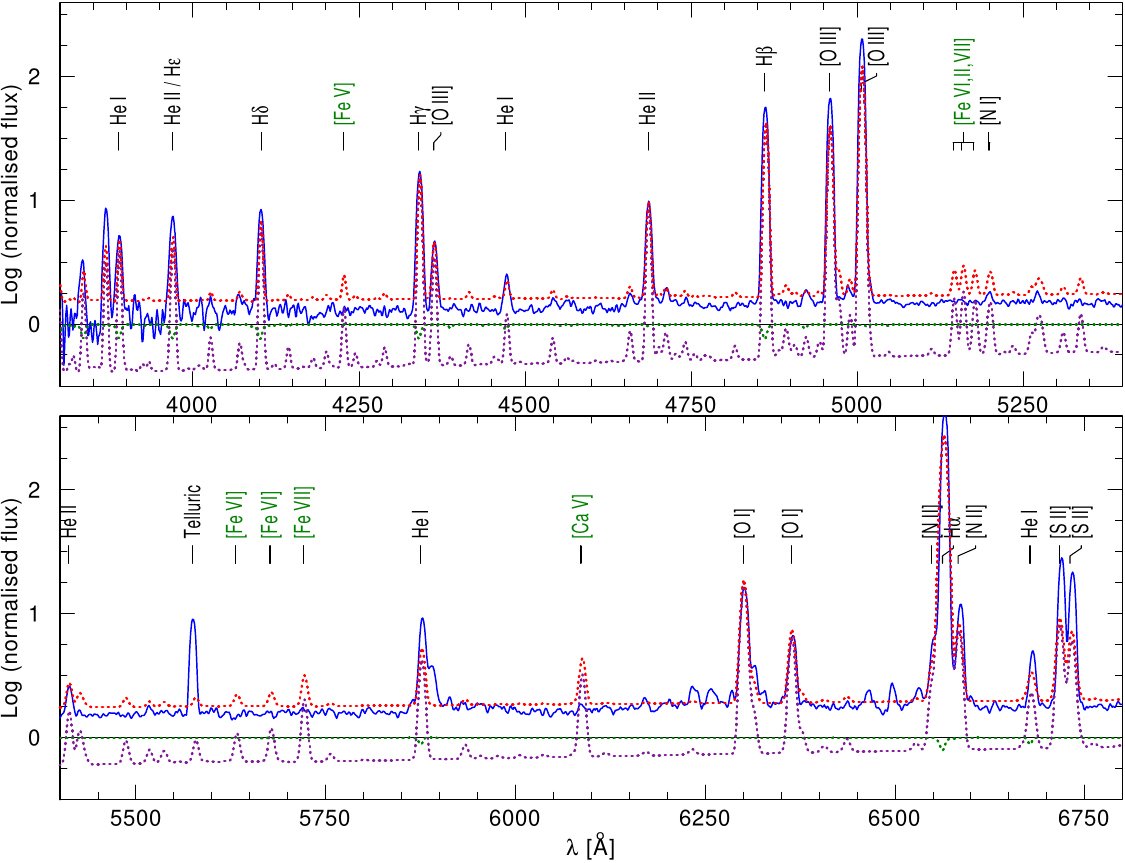}
\caption{
Optical spectrum of Ho\,II X-1 taken with the \textit{Subaru} telescope. The spectrum is normalised to the stellar model continuum and corrected for extinction.
In blue, the observations, in dotted green the synthetic stellar spectrum from the \powr\ model for a B0.5I supergiant (${T_\ast}=\,\mathrm{26\,k\,K}$, $\log g=\,3.1\,\mathrm{[cm\,s^{-2}]}$ and $\log L/L_\odot=\,5.3$), in dotted purple the \cloudy\ nebular model for an X-ray ionising source modelled with \texttt{diskir}, and in dotted red the combined \powr\ and \cloudy\ spectra.
The most relevant features are identified in black, while those identified in green correspond to high ionisation stages of iron and calcium overpredicted by the \cloudy\ model (Sect.~\ref{sec:cloudy}).
}
\label{fig:subaru+modelos}
\end{figure*}

%%%%%%%%%%%%%%%%%%%%%%%%%%%%%%%%%%%%%%%%%%%%%%%%%%%%%%%%%
%%%%% IMAGEN GTC OPTICO SPECTRUM + MODELS
%%%%%%%%%%%%%%%%%%%%%%%%%%%%%%%%%%%%%%%%%%%%%%%%%%%%%%%%%
\begin{figure*}[h!]
\centering
\includegraphics[angle=270 , width=\linewidth]{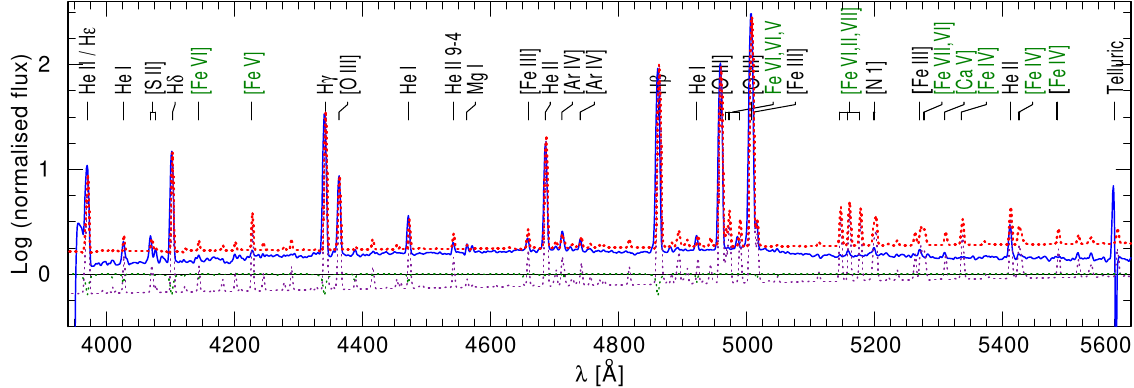}
\caption{
Same as Fig.\,\ref{fig:subaru+modelos}, but with the observed spectrum taken with the GTC.
}
\label{fig:gtc+modelos}
\end{figure*}

\section{Methods and data analysis}
%\section{Data analysis and results}
\label{sec:meth+ana}

\begin{table*}
\footnotesize\setlength{\tabcolsep}{5pt}
\centering
\setlength{\tabcolsep}{8pt}
\caption{
Parameters of the best-fit spectral models to the observed  \textit{XMM-Newton} EPIC PN spectrum of \hoii\ (ObsID:0864550501).
}\label{table:xspecmodels}
%\begin{tabular*}{\textwidth}{@{\extracolsep\fill}lccc|lccc}
\begin{tabular}{lccc|lccc}
\hline
\hline
\multicolumn{4}{c}{\rule{0cm}{2.5ex}\texttt{Tbabs$\times$(bbody+diskpbb)}} &
\multicolumn{4}{|c}{\texttt{Tbabs$\times$(diskbb+diskbb)}} \\%\cmidrule{2-4}\cmidrule{5-7}%
Component   &   Parameter	& Unit	&  Value   &   Component   &   Parameter	& Unit	&  Value \\
\hline
\rule{0cm}{3ex}\texttt{Tbabs}   &   $N_{\rm H}$   &   $10^{21}\,cm^{-2}$   &	$1.02\pm0.05$  &  
\texttt{Tbabs}   &   $N_{\rm H}$   &   $10^{21}\,cm^{-2}$   &   $0.68\pm0.04$   \\
\texttt{bbody}   &   $kT$  &   keV         &   $0.182\pm0.005$   &
\texttt{diskbb}  &   $T_{\rm in}$ &   keV         &   $0.31\pm0.01$   \\
\texttt{bbody}   &   norm    &           &	$(3.0\pm0.2)\times 10^{-5}$  &	
\texttt{diskbb}  &   norm    &           &   $29\pm4$   \\
\texttt{diskpbb} &   $T_{\rm in}$ &   keV         &   $1.95\pm0.07$   &
\texttt{diskbb}  &   $T_{\rm in}$ &   keV         &   $1.46\pm0.04$   \\
\texttt{diskpbb} &   $p$	&               &	$0.5$         &
\texttt{diskbb}  &   norm    &           &	$(4.2\pm0.5)\times 10^{-2}$   \\
\texttt{diskpbb} &   norm	&           &   $(3.9\pm0.6)\times 10^{-3}$   &
        &           &           &                   \\
\hline
\multicolumn{4}{l}{\rule{0cm}{3ex}$L_{\rm X}=(1.18\pm0.02)\times 10^{40}\,{\rm erg\,s^{-1}}$}  &
\multicolumn{4}{|l}{$L_{\rm X}=(1.02\pm0.01)\times 10^{40}\,{\rm erg\,s^{-1}}$}  \\
\multicolumn{4}{l}{\rule{0cm}{3ex}$\chi^2/{\rm d.o.f.}\;131/144$}  &
\multicolumn{4}{|l}{$\chi^2/{\rm d.o.f.}\;175/144$}  \\
\hline
\rule{0cm}{0.2cm} \\
\hline
\hline
\multicolumn{4}{c}{\rule{0cm}{3ex}\texttt{Tbabs$\times$(diskbb+comptt)}} &
\multicolumn{4}{|c}{\texttt{Tbabs$\times$redden$\times$diskir}} \\%\cmidrule{2-4}\cmidrule{5-7}%
Component   &   Parameter	& Unit	&  Value   &   Component   &   Parameter	& Unit	&  Value   \\
\hline
\rule{0cm}{3ex}\texttt{Tbabs}   &   $N_{\rm H}$      &   $10^{21}\,{\rm cm^{-2}}$   &   $1.21\pm0.01$   &
\texttt{Tbabs}   &   $N_{\rm H}$      &   $10^{21}\,{\rm cm^{-2}}$   &   $0.13\pm0.01$   \\
\texttt{diskbb}  &  $T_{\rm in}$     &	keV     &	$0.17\pm0.01$   &
\texttt{redden}  &  $E(B-V)$	&           &   0.05  \\
\texttt{diskbb}  &  norm	&           &	$329\pm104$   &
\texttt{diskir}  &  $kT_{\rm disk}$ &	keV     &	$0.18\pm0.01$   \\
\texttt{compTT}  &  redshift	&	    &   0         &
\texttt{diskir}  &  $\Gamma$   &           &	$2.16\pm0.04$   \\
\texttt{compTT}  &  $T_0$	    &    keV	&   $0.17\pm0.01$   &
\texttt{diskir}  &  $kT_{\rm e}$	&    keV	&   $2$     \\
\texttt{compTT}  &  $kT$  	&    keV	&   $2.8\pm0.8$   &
\texttt{diskir}  &  $L_{\rm c}/L_{\rm d}$	&           &	$0.83\pm0.05$   \\
\texttt{compTT}  &  $\tau_{\rm p}$  &	        &   $4.5\pm0.9$   &
\texttt{diskir}  &  $f_{\rm in}$     &           &   $0.1$   \\
\texttt{compTT}	 &  approx  &           &   $1$   &
\texttt{diskir}  &  $r_{\rm irr}$    &           &   $1.1$   \\
\texttt{compTT}  &  norm    &           &   $(1.6\pm0.3)\times 10^{-3}$   &
\texttt{diskir}  &  $f_{\rm out}$ 	&           &   $5\times 10^{-4}$   \\
        &           &           &                 &    
\texttt{diskir}  &  $\log r_{\rm out}$ &       	&   $7$   \\
        &           &           &                 &				
\texttt{diskir}	 &  norm	&           &   $324\pm88$    \\
\hline
\multicolumn{4}{l}{\rule{0cm}{3ex}$L_{\rm X}=(1.3\pm0.8)\times 10^{40}\,{\rm erg\,s^{-1}}$ } &
\multicolumn{4}{|l}{$L_{\rm X}=(1.3\pm0.7)\times 10^{40}\,{\rm erg\,s^{-1}}$}  \\
\multicolumn{4}{l}{\rule{0cm}{3ex}$\chi^2/{\rm d.o.f.}\;109/144$}  &
\multicolumn{4}{|l}{$\chi^2/{\rm d.o.f.}\;128/141$}  \\
\hline
\end{tabular}
\tablefoot{ 
The spectral fitting was done using the {\sc xspec} X-ray spectral fitting software \citep[][see detailed description of the models in the {\sc xspec} manual]{xspec}.
The ISM absorption towards \hoii\ is included using the \texttt{Tbabs} model \citep{Wilms2000}.
Four models, previously discussed in the literature to model the X-ray spectra of \hoii\, are fitted as is shown in the four quadrants in the table. In each quadrant, the first column gives the name of the spectral component of the corresponding model in {\sc xspec}, the second column gives the model parameter, the third column gives units, and the fourth column gives the value and the error of the corresponding parameter. Those values that do not have an associated error in the table were fixed during the fitting process. 
In the \texttt{Tbabs$\times$(bbody+diskpbb)} model, $p$ is fixed at 0.5 \citep{Kajava2012}. In the \texttt{Tbabs$\times$(diskbb+comptt)} model the seed temperature for the \texttt{comptt} model is fixed as being the same as the $T_{\rm in}$ in the \texttt{diskbb} model \citep{Pintore2014}. In the \texttt{Tbabs$\times$redden$\times$diskir} model, $T_{\rm e}$, $f_{\rm in}$ and $r_{\rm irr}$ are fixed during the fitting \citep{Tao2011}. $L_{\rm X}$ is calculated for a distance of 3.05\,Mpc.}
%\footnotetext[1]{Example for a first table footnote.}
%\footnotetext[2]{Example for a second table footnote.}
\end{table*}

The spectra analysed in this work are displayed in Figs.\,\ref{fig:uv+models}\,to\,\ref{fig:gtc+modelos}.
The UV spectrum of \hoii\ (Fig.\,\ref{fig:disentangling}) shows narrow absorption lines of C\,{\sc ii}\,$\lambda\lambda$\,1334,1335\,\r{A}, Si\,{\sc ii}\,$\lambda$\,1527\,\r{A}, Fe\,{\sc ii}\,$\lambda$\,1608\,\r{A}, Al\,{\sc ii}\,$\lambda$\,1671\,\r{A}\, and Al\,{\sc iii}\,$\lambda$\,1855\,\r{A}. These lines are unresolved, as their widths do not exceed the low spectral resolution of the observations.
%grating G140L centred at 1280\,\r{A} ($R\sim$2000).
They correspond to transitions from ground levels of low ionisation stage ions \citep{NIST_ASD}. On this basis, we conclude that these lines are due to the line-of-sight absorption in the interstellar medium (ISM).  
The spectrum shows strong emission lines, which are generated in the high-excitation nebula around the ULX.
These emission lines are somewhat broader than the ISM absorption lines, because the nebula emission fills the entire 2\farcs5 aperture of COS, smearing the lines and effectively reducing the spectral resolution.

The optical spectra of \hoii\ (Figs.\,\ref{fig:subaru+modelos} and \ref{fig:gtc+modelos}) are dominated by its high excitation nebula. The emission lines are unresolved, with their widths corresponding to the spectral 
resolution of the FOCAS and OSIRIS spectrographs.

To proceed with the data analysis, we construct the SED of \hoii\ by combining our UV photometric data with the measurements in the optical and near-IR from \citet{Tao2012}. The authors reported the average fluxes of non-simultaneous observations, with and without nebular subtraction, in the  F814W, F555W, F450W and F336W filters. The SED of the ULX is shown in Fig.\,\ref{fig:sed} (see Sect.\,\ref{sec:phot} for a detailed discussion of the results) and the flux values are reported in Table\,\ref{table:photflux}.

%%%%%%%%%%%%%%%%%%%%%%%%%%%%%%%%%%%%%%%%%%%%%%%%%%%%%%%%%
%%%%% PoWR
%%%%%%%%%%%%%%%%%%%%%%%%%%%%%%%%%%%%%%%%%%%%%%%%%%%%%%%%%
\subsection{Spectral modelling of the donor star}
\label{sec:powr}

Model calculations of the donor star spectrum are performed with the Potsdam Wolf-Rayet (\powr ) model atmosphere code. \powr\ is a state-of-the-art code for stellar atmosphere in spherical symmetry and with winds. Local thermodynamic equilibrium is not presumed (non-LTE) \citep{PoWRHainich2019,PoWRTodt2015,PoWRSander2015,PoWRH&G2004,PoWRGrafener2002}. The basic input parameters for a model are the luminosity ($L_\ast$), the stellar temperature ($T_\ast$), the surface gravity ($\log g$), and the mass-loss rate ($\dot M$) of the star. Also, the abundances have to be specified. 
We include C, N, O, Si, P, Al, S, Mg and the Fe-group elements, and adopt their down-scaled solar abundances according to the metallicity of 0.07 Z$_\odot$ found for the host galaxy Holmberg\,II \citep{Kaaret2004}.

As there is no clear evidence for stellar lines in the optical and only marginal evidence for such lines in the UV, the stellar parameters can only be roughly constrained.
Motivated by the work of \citet{Fabrika2015}, we first compare the observations with \powr\ models for WR stars at low-metallicity. However, they cannot reproduce either the SED or the UV and optical spectra, as they predict very broad emission lines that would be visible even with the strong nebular contribution. 

Models for OB-type stars are tested next.
The stellar temperature and luminosity can be estimated from the SED (Sect.~\ref{sec:phot}), particularly from the lack of a Balmer jump. % Additionally, the temperature range is constrained by the strength of the forest of iron lines, especially at wavelengths above 1700\,\r{A} (Fig.\,\ref{fig:uv+models}).
This restricts the temperature to within 20 to 40\,kK.
Thus, if the observed continuum belongs to the donor star in the system, it could be an early B supergiant. More constraints on the temperature are obtained from the UV spectrum. 
The lack of a blue-shifted absorption component in the C\,{\sc iv} doublet at 1548/50\,\r{A} and the He\,{\sc ii} line at 1640\,\r{A}, in addition to the tentative presence of weak absorption lines at wavelengths above 1700\,\r{A} (Fig.\,\ref{fig:uv+models}), constrain the temperature range even more, between 20 and 30\,kK.
This is also consistent with the tentative detection of C\,{\sc ii}\,$\lambda\lambda$\,1334,1335\,\r{A} and \alii . 

The surface gravity and the wind mass-loss rate are poorly constrained due to the lack of observed non-nebular spectral lines. 
To get a rough estimate of these parameters, we employ a pre-computed grid of stellar atmosphere models \citep{PoWRHainich2019}, based on the MIST stellar evolution models \citep{MISTDotter2016,MISTChoi2016,MISTPaxton2011}. 
As customary in stellar spectroscopy, we select our preferred model by comparison with the observations through multiple iterations. During this process, we systematically adjust the model's stellar parameters and conduct visual inspections. 
The selected model is the one that best reproduces the following spectral features: absence of a noticeable Balmer jump; lack of strong blue-shifted absorption components of the UV lines C\,{\sc iv}\,$\lambda\lambda\,1548,1550$\,\r{A} and He\,{\sc ii}\,$\lambda\,1640$\,\r{A}; weak absorption lines of N\,{\sc iii}\,$\lambda\lambda\,1747,1754$\,\r{A} and iron, and lack of strong absorption features.
%The selected model is the one that best fits the spectral features and SED (Figs.\,\ref{fig:uv+models} and \ref{fig:sed}).
Thus, the derived donor star parameters ($T_\ast\simeq26$\,kK, $\log L/L_\odot\simeq5.3$, $\log g\simeq3.1$\,[$\mathrm{cm\,s^{-2}}$], $\log \dot M\simeq-8.5\,$[$\mathrm{M_\odot\,yr^{-1}}$]) correspond to a star of spectral type B0.5\,I \citep{Searle2008,Martins2021}, which is a common type of a donor star in high mass X-ray binaries \citep{Martinez-Nunez2017}.

%%%%%%%%%%%%%%%%%%%%%%%%%%%%%%%%%%%%%%%%%%%%%%%%%%%%%%%%%
%%%%% XSPEC
%%%%%%%%%%%%%%%%%%%%%%%%%%%%%%%%%%%%%%%%%%%%%%%%%%%%%%%%%
\subsection{X-ray spectral modelling}
\label{sec:xray}

The EPIC-PN spectrum of \hoii\ is analysed using the {\sc xspec} spectral fitting package \citep{xspec}.
Four different models were tested. 
The irradiated disc model, \texttt{diskir}  \citep{Gierlinski2009-diskir},  
can reproduce the UV and optical photometric measurements of \hoii\ \citep{Tao2012}.
Another model is a composite of a blackbody, \texttt{bbody}, and a slim disc, \texttt{diskpbb} \citep{Mineshige1994-diskpbb}, commonly used in the analysis of the ULX X-ray spectra  \citep{Kajava2012}. 
The next model combines two multi-temperature blackbody discs, \texttt{diskbb} \citep{Mitsuda1984-diskbb,Gurpide2021A}. 
Finally,  we test a multicolour blackbody disc together with a Comptonising component, \texttt{comptt} \citep{Titarchuk1994-comptt,Pintore2014}.
The best-fit parameters of all these models are given in Table~\ref{table:xspecmodels}. 
We note that the \texttt{diskir} model might not be fully physically motivated in the case of \hoii , as the parameters $f_{\rm out}$ and $\log r_{\rm out}$ are unconstrained. %it does well extrapolate the flux from the X-rays to the UV and optical, and is the best model to estimate the ionising flux in the extreme UV.
However, this model reproduces fluxes in the UV and optical wavelength
range, making it a suitable choice for extrapolating the ionising spectrum from X-ray to UV and optical wavelengths.
The other three models do not reproduce the photometric data in the UV and optical ranges. Therefore, they need to be combined with our donor star model (B0.5\,I star with $T_\ast=26$\,kK and $\log L/L_\odot=5.3$) to explain the UV and optical measurements. In this way, we construct four alternative SEDs as ionising spectra input for the nebula models.

%%%%%%%%%%%%%%%%%%%%%%%%%%%%%%%%%%%%%%%%%%%%%%%%%%%%%%%%%
%%%%% CLOUDY
%%%%%%%%%%%%%%%%%%%%%%%%%%%%%%%%%%%%%%%%%%%%%%%%%%%%%%%%%
\subsection{Spectral modelling of the Foot Nebula}
\label{sec:cloudy}

To obtain a synthetic spectrum of the nebular emission, we used the photoionising code \cloudy\ \citep{CLOUDY2017}.
The key input parameters for the nebula model are the luminosity and the SED of the ionising source; that is, the broadband X-ray and UV (XUV) flux and spectrum of \hoii .
The bolometric luminosity of the source is fixed at its average value of $L_{\rm bol}=1.4\times 10^{40}$\,erg\,s$^{-1}$ \citep{Grise2010,Berghea2010I}. While \hoii\ has strong X-ray variability on a timescale of days \citep{Gurpide2021B}, this approach is reasonable as the recombination timescale of the nebula is on the order of thousands of years. 

To produce the broadband XUV SED, we extrapolate the spectral models from X-rays to UV (Sect.~\ref{sec:xray}). 
While the source is well observed in X-rays and UV, 
the extreme UV wavelength range that is at the ionisation threshold of He\,{\sc ii} (around $\lambda$\,227.84\,\r{A}) cannot be observed. Therefore, the spectral shape at these wavelengths is model-dependent.
We test four alternative SEDs obtained when fitting the X-ray spectrum (Sect.~\ref{sec:xray}). As they are all normalised to the same X-ray luminosity, the main difference between them is the contribution to the extreme UV range predicted by the different disc models.
For each of the four SEDs, a grid of \cloudy\ models is computed by varying the density and the inner radius of the He\,{\sc iii} nebula. 

Further input parameters to the \cloudy\ model include the hydrogen density, the inner radius of the nebula, and the nebular chemical abundances.
The electron density and temperature of nebulae are typically constrained using flux ratios between certain forbidden metal lines, such as [S\,{\sc ii}]\,$\lambda\lambda\,6716,6731\,$\r{A} and  [O\,{\sc iii}]\,$\lambda\lambda\,4363,4959,5007\,$\r{A} \citep{elosterbrock2006agna.book.....O}. Fortunately, the Subaru spectra cover these important diagnostic lines. Using these methods we roughly constrain the electron temperature to $\mathrm{14.5\,kK}$ and the electron density to $\mathrm{125\,cm^{-3}}$.  
These values are used as input for an initial \cloudy\ model. However, the agreement with the observed spectrum is not perfect. 
Likely, this is because the [O\,{\sc ii}]$\lambda\lambda\,3426,3729,7319,7330$\,\r{A} auroral lines, which are needed to better constrain the electron density, are outside of the wavelength range of the available spectra. 
To improve the quality of our models, we compute a grid of models for densities between 10 to 200\,cm$^{-3}$ with steps of 10\,cm$^{-3}$. This procedure is repeated for each of the four SEDs.  

For the geometry of the nebula, we choose spherical symmetry. 
The inner radius is a free parameter of our \cloudy\ models. 
In the \hst\ image shown in the insert in Fig.\,\ref{fig:hstneb} the nebula around the point source does not have a ring-like structure. Thus, taking into account the spatial resolution of the telescope, the maximum value of the inner radius must be smaller than 1\,pc ($\approx 0\farcs07$ at \hoii\ distance).
We compute models for inner radii between 10$^{-5}$ to 1\,pc and conclude that the effect on the synthetic spectrum is negligible.

Finally, the nebular abundances of oxygen, nitrogen, sulfur, neon, and argon are adopted from \cite{Croxall2009}. They obtained these values after performing a spectroscopic analysis of the H\,{\sc ii} region around \hoii . 
%where the H\,{\sc ii} region around \hoii\ was analysed. 
The abundances of other elements are scaled down from solar values \citep{Asplund2009} to 0.07Z$_\odot$. 
Although \citet{Egorov2013} reported a higher mean metallicity for the host galaxy, Holmberg\,II, these authors also adopted the values from \citet{Croxall2009} for the region containing the Foot Nebula. 

Following these criteria, we computed a grid of nebula models. To identify models suitable for comparison with the observations, we first examine the size of the He\,{\sc iii} region surrounding the ULX Ho\,II. The observed radius of $\sim15$\,pc \citep[][Fig.\,\ref{fig:hstneb}]{Kaaret2004} is compared with the predicted sizes from our models. 
Given the model SED, this procedure constrains the electron density of the gas.
%The radius of the He\,{\sc iii} region surrounding the ULX Ho\,II X-1 is  $\sim15$\,pc \citep{Kaaret2004} (Fig.\,\ref{fig:hstneb}). 
%Given the model SED, the size of this region constrains the electron density of the gas. 
For the \texttt{diskir} and \texttt{diskbb+diskbb} models (Sect.~\ref{sec:xray} and Table~\ref{table:xspecmodels}), the size of the He\,{\sc iii} region is reproduced when the gas density is rather low, $n_{\rm e}=20\pm 10$\,cm$^{-3}$, in rough agreement with \citep{Kaaret2004}.  
On the other hand, the \texttt{diskbb+comptt} and  \texttt{bbody+diskpbb} models (Sect.~\ref{sec:xray} and Table~\ref{table:xspecmodels}), favour higher electron densities, $n_{\rm e}=140\pm 10$\,cm$^{-3}$  and $50\pm 10$ cm$^{-3}$, respectively.  
To proceed, we compute a grid of synthetic spectra combining models for X-rays, the donor star, and the nebula, which could be directly compared with the observations.

%%%%%%%%%%%%%%%%%%%%%%%%%%%%%%%%%%%%%%%%%%%%%%%%%%%%%%%%%
%%%%%%%%%%%%%%%%%%%%%%%%%%%%%%%%%%%%%%%%%%%%%%%%%%%%%%%%%
%%%%%%%%%%%%%%%%%%%%%%%%%%%%%%%%%%%%%%%%%%%%%%%%%%%%%%%%%
%%%%% R E S U L T S
%%%%%%%%%%%%%%%%%%%%%%%%%%%%%%%%%%%%%%%%%%%%%%%%%%%%%%%%%
%%%%%%%%%%%%%%%%%%%%%%%%%%%%%%%%%%%%%%%%%%%%%%%%%%%%%%%%%
%%%%%%%%%%%%%%%%%%%%%%%%%%%%%%%%%%%%%%%%%%%%%%%%%%%%%%%%%

\section{Results}
\label{sec:results}

%%%%%%%%%%%%%%%%%%%%%%%%%%%%%%%%%%%%%%%%%%%%%%%%%%%%%%%%%
%%%%% IMAGEN SED
%%%%%%%%%%%%%%%%%%%%%%%%%%%%%%%%%%%%%%%%%%%%%%%%%%%%%%%%%
\begin{figure*}[h!]
\centering
\includegraphics[width=\textwidth]{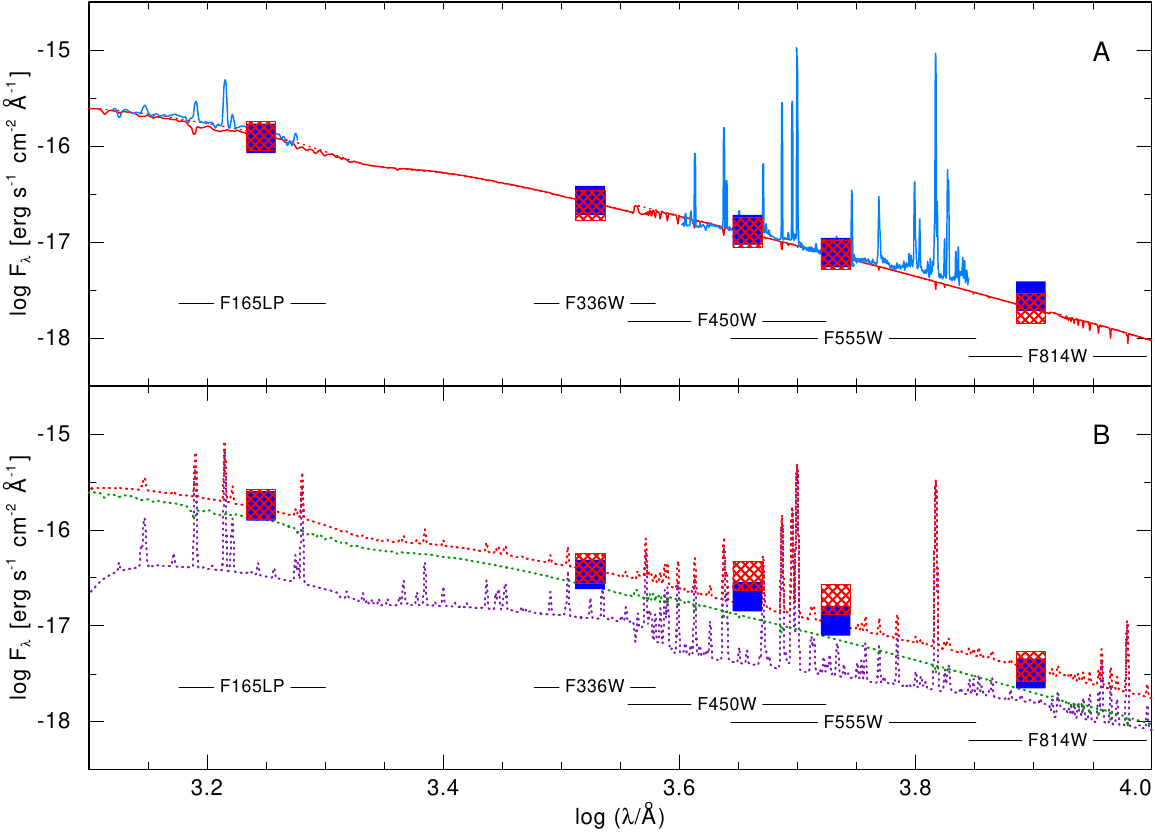}
\caption{
Spectral energy distribution (SED) of the ULX \hoii\ from UV to optical. 
{\it Upper panel A}:
The blue line shows the data extracted from the observations with \hst\ COS (UV), GTC OSIRIS and Subaru FOCAS (both in the optical). The blue boxes display photometric measurements from the \hst\ ACS SBC F165LP in the UV and from \hst\ WFPC2 in the optical (Sect.~\ref{sec:phot}).
The black lines at the bottom represent the filter wavelength ranges, each of them identified by the name of the filter. 
The red line indicates the synthetic \powr\ SED for a B0.5I supergiant with ${T_\ast}=\,\mathrm{26\,k\,K}$, $\log g=\,3.1\,\mathrm{cm\,s^{-2}}$ and $\log L/L_\odot=\,5.3$. The red hatched boxes show the modelled photometry.
{\it Bottom panel B}:
The blue boxes are the photometric measurements of the \hoii\ counterpart from \hst\ filters without any correction for the nebular contribution.
The purple line shows the synthetic \cloudy\ SED for a nebula photoionised by an X-ray source, as is described by a \texttt{diskir} model (Sect.~\ref{sec:xray}, Table~\ref{table:xspecmodels}), and with a luminosity of $L_{\rm bol}=\,1.4\times 10^{40}\,$erg\,s$^{-1}$.
The green line represents the synthetic \powr\ SED for a B0.5I supergiant with ${T_\ast}=\,\mathrm{26\,k\,K}$, $\log g=\,3.1\,\mathrm{[cm\,s^{-2}]}$ and $\log L/L_\odot=\,5.3$.
The red line is the combination of both \powr\ and \cloudy\ models, and the red-hatched boxes represent the modelled photometry for the combined model. 
}
\label{fig:sed}
\end{figure*}

%-------------------
%-------------------

\subsection{Photometry and spectral energy distribution}
\label{sec:phot}

Before comparing the synthetic stellar and nebular SEDs with the photometric data, we correct the models for extinction, taking into account both the Galactic and the Holmberg\,II contributions. 
The synthetic SED of the donor star matches the photometry in the UV and the optical (upper panel of Fig\,\ref{fig:sed}). Furthermore, the combined stellar and nebular model also matches the photometric measurements that also comprise the nebular contribution
(bottom panel of Fig.\,\ref{fig:sed}).
At this point, it is necessary to point out that no optical excess is observed as would be expected if the system were accreting supercritically. During super Eddington accretion, the disc must contribute to the total luminosity significantly, producing a measurable bump in the optical wavelength range \citep{Ambrosi2018}.

In the IR wavelength range, an excess in the SED of \hoii\ has been reported by \citet{Lau2017,Lau2019} and attributed to a circumbinary disc of heated dust around the donor star. The IR excees could also indicate that the system is undergoing a rapid phase of Roche Lobe overflow that fuels the accretion process and forms a circumbinary disc where dust formation occurs. 
Although the dust emission might, in principle, contaminate the redmost data we consider here (see the \hst\ WFPC2 filter F814W in Fig.\,\ref{fig:sed}), our models indicate that this is not the case, since the combined (stellar plus nebular) SED contribution matches the photometry very well (lower panel in Fig.\,\ref{fig:sed}).

%--------------------------
%--------------------------

\subsection{Ultraviolet and optical spectra of \hoii }
\label{sec:uvlines}

\begin{table*}
\footnotesize\setlength{\tabcolsep}{5pt}
\centering
\setlength{\tabcolsep}{8pt}
\caption{Stellar parameters for the donor star in the \hoii\ obtained from spectroscopic analysis}
\label{table:powrmodels}
\begin{tabular}{lccc}
\hline
\hline
%\multicolumn{4}{c}{\rule{0cm}{3ex}\powr\ stellar model} \\
\rule{0cm}{2.5ex}
Parameter   &   Considered range  &   Valid range  &   Favourite Model  \\
\midrule
$\log L/L_\odot$  &   $\numrange{4.1}{5.4}$   &   $\numrange{5.2}{5.3}$   &   $5.3$   \\
${T_{\ast}}$ $[{\rm k\,K}]$    &   $\numrange{15}{110}$   &   $\numrange{20}{30}$   &   $26$   \\
$\log g$ $[{\rm cm\,s^{-2}}]$   &   $\numrange{3.0}{3.6}$   &   $\numrange{3.0}{3.4}$   &   $3.1$   \\
$\log \dot M$ $[\mathrm{M_\odot\,yr^{-1}}]$   &    $\numrange{-9.1}{-7.7}$ &   $\lesssim-8.0$   &   $-8.5$   \\
\hline
\end{tabular}
\end{table*}

Even though the UV spectrum of \hoii\ seems to be completely dominated by interstellar absorption and nebular lines, when it is compared with our combined stellar plus nebula model one can see tentative evidence of stellar contribution (see Fig.\,\ref{fig:uv+models}). It cannot be ruled out that the C\,{\sc ii} doublet at 1334/1335\,\r{A} is blended with the Si\,{\sc iii}\,$\lambda\lambda$\,1341,1342,1343\,\r{A} triplet, which, according to our selected donor star model, is present in the stellar atmosphere. Furthermore, the profile of the nebular O\,{\sc iv}\,$\lambda\lambda$\,1401,1404\,\r{A} lines could be influenced by stellar contribution from Si\,{\sc iv}\,$\lambda\lambda$\,1394,1402\,\r{A}.
Importantly, the \powr\ model does not predict any other stellar feature that would be detectable in the observations, given the limited S/N and the blending with strong nebular emission.

Previous analysis of optical spectra from \hoii\ claimed the detection of broad emission lines He\,{\sc ii}\,$\lambda$\,4686\,\r{A}, H$\alpha$ and H$\beta$ \citep{Fabrika2015}. On this basis, the optical spectrum was misclassified as a WR-type.
Based on the presence of He\,{\sc ii} and H lines but the absence of He\,{\sc i} lines, it was suggested that the optical counterpart of the ULX originates from the accretion disc.  
This was presented as proof that the compact object in \hoii\ accretes supercritically (i.e. at a rate highly above the Eddington limit). 
However, our observations with two times higher spectral resolution and our quantitative analysis show that the line width was mistakenly attributed to an expanding wind. 
Instead, the line broadening is fully explained by instrumental broadening and is in accordance with the low spectral resolution of the \textit{Subaru} FOCAS. These lines are much narrower in the higher-resolution spectra obtained with the GTC OSIRIS.
Hence, our independent analysis does not confirm the presence of disc-wind lines (Figs.\,\ref{fig:subaru+modelos}, \ref{fig:gtc+modelos}), removing another reason for suggesting accretion in a supercritical regime.

Standard accretion (or decretion) discs often display emission lines, such as H$\alpha$, H$\beta$ or He\,{\sc ii}\,$\lambda\,4686\,$\r{A} and the Bowen blend,
in their optical spectra. Depending on inclination, the lines might have a characteristic double peak profile as indeed observed in the optical spectra of black hole LMXBs  \citep{Soria2000,Panizo-Espinar2022}. The emission lines that could be attributed to the accretion disc are not detected in the optical spectrum of \hoii. However, the optical spectra of \hoii\ are quite noisy and strongly dominated by the nebula lines and the donor star continuum. Therefore, any possible spectral feature of the accretion disc is most likely strongly diluted and not detectable in our data.  
Interestingly, the disc lines are also not detected in the optical spectra of the prominent black hole HMXB  M33 X-7, because the radiation of the OB donor star in the optical and UV is significantly brighter compared to the possible disc contribution \citep{Ramachandran2022}.

To perform the combined analysis of the UV and optical data of \hoii\ and its nebula, we first scale the synthetic \cloudy\ spectra to the different aperture (COS) and slit sizes of the spectrographs (FOCAS \& OSIRIS).
%As a first step, to enable the comparison, the synthetic \cloudy\ spectra are scaled correspondingly to the different aperture (COS) and slits sizes of spectrographs (FOCAS \& OSIRIS). 
Then, the model flux in the H\,$\beta$ line and the ratios between fluxes in metal lines and in H\,$\beta$ are compared with those measured from the observed spectra. 
The {\sc xspec}+\cloudy\ model that reproduces the line ratios best has the SED described by the \texttt{diskir} model and the nebular electron density of $n_{\rm e}=20\pm 10$ cm$^{-3}$. 
Combining this model with the \powr\ stellar model for the donor star allows us, for the first time, to consistently reproduce the multi-wavelength, UV and optical, spectrum of Ho\,II X-1 (Figs.\,\ref{fig:uv+models},~\ref{fig:subaru+modelos},~\ref{fig:gtc+modelos}).

Still, there are some small deviations between the modelled and observed optical spectra. All synthetic spectra overpredict the line strengths of the high ionisation stages of iron, and underpredict the lines that are likely to be the result of shock ionisation, such as [S\,{\sc ii}]$\lambda\lambda\,6716,6731\,$\r{A}. We attribute these mild discrepancies  to 
the deviation from the spherical symmetry, probable inhomogeneities, and other contributions to the ionisation such as the shocks produced by jets or the possible contribution to the nebula from the bow-shocked gas caused by \hoii\ moving away from its natal cluster \citep{Egorov2017}.

%%%%%%%%%%%%%%%%%%%%%%%%%%%%%%%%%%%%%%%%%%%%%%%%%%%%%%%%%
%%%%%%%%%%%%%%%%%%%%%%%%%%%%%%%%%%%%%%%%%%%%%%%%%%%%%%%%%
%%%%%%%%%%%%%%%%%%%%%%%%%%%%%%%%%%%%%%%%%%%%%%%%%%%%%%%%%
%%%%% D I S C U S S I O N
%%%%%%%%%%%%%%%%%%%%%%%%%%%%%%%%%%%%%%%%%%%%%%%%%%%%%%%%%
%%%%%%%%%%%%%%%%%%%%%%%%%%%%%%%%%%%%%%%%%%%%%%%%%%%%%%%%%
%%%%%%%%%%%%%%%%%%%%%%%%%%%%%%%%%%%%%%%%%%%%%%%%%%%%%%%%%

\section{Discussion}
\label{sec:disc}

The first consistent analysis of the X-ray, UV and optical spectra of \hoii\ and its nebula presented in this paper successfully identifies the donor star and reproduces the observations without invoking super-Eddington accretion. Indeed, spectral features associated with powerful and fast outflows, as predicted and observed in supercritically accreting X-ray binaries, are absent in the UV and optical as well as in the X-ray spectra \citep{Barra2023} of \hoii\ at the moment of our observations. 
This gives important information about the nature of the compact object in \hoii , suggesting that it is a massive black hole.

\subsection{New estimates of the black hole mass and accretion rate in \hoii }

The results presented in this paper suggest that \hoii\ consists of a B supergiant donor star and a black hole accreting at or below the Eddington limit. This allows us to use the X-ray luminosity for estimating the black hole mass. 

The X-ray emission of \hoii\ is variable but its spectral appearance is persistent and typical 
%of a ULX as indeed expected
for sources accreting close to the Eddington limit \citep{Kaaret2017,King2023}. 
The time-averaged X-ray luminosity is best estimated from nebula analysis.
To reproduce the  UV and optical spectra of the He\,{\sc iii} region around \hoii , an X-ray luminosity of $L_X\sim$1$\times $10$^{40}\,{\rm erg\,s^{-1}}$ is required, which is in excellent agreement with the actual X-ray luminosity during our \hst\ COS observations (see Table\,\ref{table:xspecmodels}). 
Assuming spherical symmetry and electron scattering as the principal source of opacity, the Eddington luminosity of a black hole is given by 
$L_{\rm Edd}\approx 2.5\times 10^{38} (1+X)^{-1} M/M_\odot$\,erg\,s$^{-1}$, where $X=0.65$ is the typical hydrogen mass fraction of the accreted matter; that is, the donor star atmosphere \citep{vanParadijs1984,Poutanen2007}.
Using the X-ray luminosity during \textit{XMM-Newton} observations conducted simultaneously with our \hst\ spectroscopic observations (Table~\ref{table:xspecmodels}), the mass of the black hole in \hoii\ is  $M_\bullet  \gtrsim66\,M_\odot$.
The X-ray luminosity, and this the implied black hole mass, might be overestimated if the radiation is beamed towards us. However, our analysis shows no evidence of beaming on \hoii , as the X-ray luminosity estimated from the nebula analysis and the X-ray observations agree. Additionally, there are no indications of beaming in the existing literature for this ULX.

The accretion rate can be inferred from a thin disc model with an inner temperature constrained by the X-ray modelling. 
Following \citet{Shakura1973}, \hoii\ is accreting with $\dot M_{\rm acc} \mathrm{\sim1.8\times 10^{-6}\,}M\mathrm{_\odot yr^{-1}}$; that is, at a rate that is marginally below the Eddington limit.

\subsection{Previous works suggesting that \hoii\ is powered by a black hole}

The nature of the compact object in \hoii\ has been extensively discussed in the literature with many authors agreeing that the compact object in this ULX is a black hole. 
One of the strongest evidence for this hypothesis is the presence of a collimated jet \citep{Cseh2014,Cseh2015}. The large kinematic energy carried away by the jet implies a minimum black hole mass of $M_\bullet >25$\,\msun . This agrees with previous analysis of the source based on its X-ray luminosity \citep{Grise2010}.

More detailed studies of the X-ray spectra of \hoii\ have been performed to better constrain the mass of the compact object and the accretion rate. As ULXs are variable X-ray sources, it is possible to analyse the changes in the relationship between the accretion disc luminosity and temperature. In the standard thin accretion disc model \citep{Shakura1973}, the disc luminosity and temperature are related as $L\propto T^4$. If confirmed from the analysis of X-ray spectra, the presence of such correlation provides strong support to the assumption that the luminosity of a source is  $< 0.3 L_{\rm Edd}$. Such analysis has been conducted for \hoii. \cite{Feng2009} concluded that \hoii\ might be accreting below the Eddington limit, but they stressed the necessity of repeating the analysis with a larger spectral sample. This is achieved in \citet{Barra2023} who analysed 17 {\it XMM-Newton} observations of \hoii\ spanning over $\approx 20$\,yr. They fit the X-ray spectra using a spectral model that combines two modified black body components \citep[see details in][]{Barra2023} and found that the luminosity of the lower temperature component does correlate with $T^4$. Still, the correlation breaks for the hotter component when the bolometric luminosity exceeds $5\times 10^{39}{\rm \,erg\,s^{-1}}$. They consider this value as the $L_{\rm Edd}$  in \hoii, and, on this basis, estimate  $M_\bullet\approx 36\,M_\odot$.

Unfortunately, there are caveats in this method because the disc temperature is deduced from the fitting of X-ray spectra using a handful of existing ready-to-use models, and hence is highly model-dependent. When considering the same data as analysed in \citet{Barra2023} we notice that some observations are mildly piled up, which may affect the spectral fitting results. Furthermore, the temperatures derived from spectral analysis depend on the spectral model describing interstellar absorption.
For example, for the observations ObsID 0864550501 (simultaneous with the \hst\ observations) we determine $L_X$(0.2-12.0\,keV)$\approx 10^{40}$\,erg\,s$^{-1}$, which is twice higher than the one reported in \citet{Barra2023}.  Taking at face value, the estimated black hole mass is  $M_\bullet\approx 70\,M_\odot$. Interestingly, \citet{Barra2023} report the largest deviation from the $L\propto T^4$ relation 
for the observation ObsID 0724810101. For the same data we derive $N_{\rm H}=(8.4 \pm 1.2)\times 10^{21}$\,cm$^{-2}$ using the standard absorption model \citep{Morrison1983} and lower temperatures, $kT_1=0.29\pm 0.05$\,keV and $kT_2=1.86\pm 0.07$\,keV ($\chi^2=108$ for 78 d.o.f) compared to \citet{Barra2023}. 

An upper limit for the compact object was proposed by \citet{Goad2006} by comparing the X-ray properties of \hoii\ with that of the microquasar GRS 1915+105. The authors concluded that the black hole in \hoii\ should be more massive than a stellar compact object, but less massive than 100\,\msun .
Clearly, a better understanding of circumstellar gas around \hoii\ and more physically motivated disc models are needed to test whether  \hoii\ sometimes accretes supercritcally.

Further evidence for a massive black hole in \hoii\ was presented by \citet{Ambrosi2022} who compared binary stellar evolution models with photometric measurements and obtained $M_\bullet \sim50$\,\msun.

\subsection{The donor star and a possible evolution scenario for \hoii }

\hoii\ is located in a star-forming region of the Holmberg\,II galaxy (Fig.\,\ref{fig:hstneb}) right next to a young, 4\,Myr old,  OB star cluster, suggesting that it
has escaped from that cluster and has a similar age.  \citep{Stewart2000,Egorov2017}.  This excludes an evolved low-mass star as the donor. The donor star in \hoii\ can only be a young, massive star. 
The absence of broad emission lines in the optical and UV spectra rules out that the companion is a WR star, while the photometry across UV, optical, and infrared wavelengths is not consistent with the donor star being a red supergiant either \citep{Lau2017}. Instead, all measurements are consistent with the donor being an OB-type supergiant \citep{Tao2011,Lau2017, Vinokurov2022,Ambrosi2022}.

Our new detailed spectroscopic analysis suggests that the donor star has B0.5I spectral type, and hence is only a few million years old. The donor star parameters are listed in Table\,\ref{table:powrmodels}. 
Unfortunately, the nebular contamination of the donor star spectra prevents any measurement of radial velocity variations, and hence impedes a direct constrain on the binary's period. 
To estimate the period, we relied on the current radius and mass of the donor (22\,$R_\odot$ and 22\,\msun, respectively), and the estimated black hole mass ($M_\bullet\gtrsim66$\,\msun). 
Making the plausible assumption of the donor filling its Roche lobe and transferring mass onto the black hole and following the approximation for the Roche lobe radius from \citet{Eggleton1983}, the binary period is $P_{\rm orb}\sim6\,{\rm d}$.

A 4\,Myr old binary system in which the primary has evolved to a 66\,\msun\ black hole, and the secondary fills its Roche lobe, should have special properties. Initially, the  primary star should be more massive than 
150\,\msun, while the secondary should have a mass of $\sim 25$\,\msun\ \citep{Ambrosi2022}. 
The binary must have initially been in a wide orbital configuration with $P_{\rm orb}\sim\SIrange{e2}{e3}{d}$, which is wide enough for the primary to develop a massive core. However, the orbital separation should be small enough to allow the system to enter a common envelope phase at later evolutionary stages. We speculate that during the common envelope phase, the secondary (current donor) does not accrete any material and the primary directly collapses into a black hole. The secondary continues its evolution until it starts filling its Roche lobe and the system enters the current ULX phase. 

During the accretion phase, the black hole accretes at the Eddington limit with $\dot M_{\rm acc} \mathrm{\sim1.8\times 10^{-6}\,}M\mathrm{_\odot yr^{-1}}$. Assuming that the mass-transfer phase lasts less than $10^{6}\,$years, the black hole will accrete less than 2\,\msun\ during this phase. This implies that the spin of the black hole is not expected to change significantly during the mass-transfer phase \citep{Fragos2015}. 

It is worth noting that during the mass-transfer phase, the surface abundances of H, He, C, N, and O of the donor star are expected to change \citep{Marchant2017}, as is indeed commonly found in the blue supergiant donor stars in high-mass X-ray binaries \citep{Hainich2020}.
That is, the surface hydrogen abundance drops to values of about $X_\mathrm{H}\approx0.65$ and the C, N, and O abundances change to the CNO equilibrium values. For consistency, we tested if our stellar atmosphere model is also consistent with these values and no discrepancies can be found. This is supported by the enhanced nitrogen abundance deduced from the high-resolution X-ray spectra of \hoii\ \citep{Barra2023}.

Eventually, the donor star will lose about half of its mass during the mass-transfer event, leaving behind a helium star of about 10-15\,\msun\ that will collapse into a black hole or a neutron star. Since the mass ratio of the system is always $q= M_\bullet/M_\ast\gg 1$ the orbit of the system is expected to further shrink during the mass-transfer, bringing the components closer together. Their masses and proximity make this system a potential progenitor of a GW event, and might provide the missing link to explain the systems with extreme mass ratios as observed with LIGO/Virgo \citep{LIGO2020}.

%%%%%%%%%%%%%%%%%%%%%%%%%%%%%%%%%%%%%%%%%%%%%%%%%%%%%%%%%
%%%%%%%%%%%%%%%%%%%%%%%%%%%%%%%%%%%%%%%%%%%%%%%%%%%%%%%%%
%%%%%%%%%%%%%%%%%%%%%%%%%%%%%%%%%%%%%%%%%%%%%%%%%%%%%%%%%
%%%%% C O N C L U S I O N S
%%%%%%%%%%%%%%%%%%%%%%%%%%%%%%%%%%%%%%%%%%%%%%%%%%%%%%%%%
%%%%%%%%%%%%%%%%%%%%%%%%%%%%%%%%%%%%%%%%%%%%%%%%%%%%%%%%%
%%%%%%%%%%%%%%%%%%%%%%%%%%%%%%%%%%%%%%%%%%%%%%%%%%%%%%%%%

\section{Summary and conclusions}
\label{sec:conc}

We present the first consistent multi-wavelength analysis of the ULX \hoii\ and its surrounding nebula. For this work, we have gathered archival optical and X-ray data, and presented the first medium-resolution UV spectrum obtained with the \hst. 

The detailed spectroscopic analysis reveals the lack of signatures of strong outflows in the UV spectra, contrary to the expectations for a supercritically accreting system. The absence of strong outflows is further corroborated by the lack of broad emission lines in the optical spectra, which also refutes claims of strong disc winds in \hoii. 

We successfully reproduce the composite spectrum of \hoii\ and the surrounding He\,{\sc iii} nebula. Using a combination of stellar atmosphere and photoionising codes, we consistently explain the observations and simultaneously characterise the nebula and its ionising source. 

Furthermore, by analysing the SED and modelling the UV spectra, we obtain compelling evidence that the donor is a B supergiant with an estimated mass of $M_\ast\sim22$\,\msun . 

Based on the nebula modelling, we estimated the averaged X-ray luminosity of the ULX, and thus constrained the mass of the compact object as $M_\bullet \gtrsim66$\,\msun. This aligns with previous works suggesting the presence of a heavy black hole in \hoii.

We conclude that the ULX \hoii\ is a close binary system consisting of a black hole with $M_\bullet \gtrsim66\,M_\odot$ and a B-type supergiant with $M_\ast\approx 22\,M_\odot$ filling its Roche lobe. Black holes within this mass range have so far only been observed via gravitational wave events \citep{LIGO2020}. 
This work not only enhances our understanding of ULXs but also contributes to the broader knowledge of massive black hole formation and evolution in binary systems.

%%%%%%%%%%%%%%%%%%%%%%%%%%%%%%%%%%%%%%%%%%%%%%%%%%%%%%%%%
%%%%%%%%%%%%%%%%%%%%%%%%%%%%%%%%%%%%%%%%%%%%%%%%%%%%%%%%%
%%%%%%%%%%%%%%%%%%%%%%%%%%%%%%%%%%%%%%%%%%%%%%%%%%%%%%%%%
%%%%%%%%%%%%%%%%%%%%%%%%%%%%%%%%%%%%%%%%%%%%%%%%%%%%%%%%%
%%%%%%%%%%%%%%%%%%%%%%%%%%%%%%%%%%%%%%%%%%%%%%%%%%%%%%%%%
%%%%%%%%%%%%%%%%%%%%%%%%%%%%%%%%%%%%%%%%%%%%%%%%%%%%%%%%%
%%%%%%%%%%%%%%%%%%%%%%%%%%%%%%%%%%%%%%%%%%%%%%%%%%%%%%%%%

\begin{acknowledgements}
We thank the referee for their useful comments.
Based on observations with the NASA/ESA Hubble Space Telescope obtained at the Space Telescope Science Institute, which is operated by the Association of Universities for Research in Astronomy, Incorporated, under NASA contract NAS5-26555. Support for Programme number HST-GO-16182 was provided through a grant from the STScI under NASA contract NAS5-26555.
Based on data from the GTC Archive at CAB (CSIC -INTA). The GTC Archive is part of the Spanish Virtual Observatory project funded by MCIN/AEI/10.13039/501100011033 through grant PID2020-112949GB-I00.
S.R.S. and D.P. acknowledge financial support by the Deutsches Zentrum für Luft und Raumfahrt (DLR) grants FKZ 50OR2108 and 50OR2005.
AACS and VR acknowledge support by the Deutsche Forschungsgemeinschaft (DFG, German Research Foundation) in the form of an Emmy Noether Research Group -- Project-ID 445674056 (SA4064/1-1, PI Sander). 
AACS and VR further acknowledge support from the Federal Ministry of Education and Research (BMBF) and the Baden-Württemberg Ministry of Science as part of the Excellence Strategy of the German Federal and State Governments.
TB was supported by the NCN grant n 2023/49/B/ST9/02777.
The collaboration of co-authors was facilitated by support from the International Space Science Institute (ISSI, Bern).
\end{acknowledgements}

%-------------------------------------------------------------------

\bibliography{references.bib}
\bibliographystyle{aa}

\begin{appendix}

\section{Observation log tables}

%%%%%%%%%%%%%%%%%%%%%%%%%%%%%%%%%%%%%%%%%%%%%%%%%%%%%%%%%
%%%%% TABLA DATOS COS
%%%%%%%%%%%%%%%%%%%%%%%%%%%%%%%%%%%%%%%%%%%%%%%%%%%%%%%%%
\begin{table}[h]
\caption{\label{table:fitsheader_COS} Observation log for the \hst -COS data. Programme ID: 16182, PI: L. Oskinova}
\centering
\begin{tabular}{lcccc}
\hline\hline
Rootname & Visit &Observation date & Observation time &Exposure time\\
 &  & UT (yyyy-mm-dd)& UT (hh:mm:ss) & (s)\\
\hline
lee701dbq & visit 1 & 2021-02-08 & 07:58:07 & 517.0 \\
lee701dfq & visit 1 & 2021-02-08 & 08:08:39 & 517.0 \\
lee701djq & visit 1 & 2021-02-08 & 08:46:12 & 517.0 \\
lee701dmq & visit 1 & 2021-02-08 & 08:56:44 & 517.0 \\
lee702l0q & visit 2 & 2021-03-17 & 05:13:32 & 2416.1 \\
lee702lbq & visit 2 & 2021-03-17 & 06:48:44 & 3069.1 \\
lee702lkq & visit 2 & 2021-03-17 & 09:59:12 & 3069.1 \\
lee703pnq & visit 3 & 2021-03-18 & 05:01:56 & 2416.1 \\
lee703pvq & visit 3 & 2021-03-18 & 06:37:08 & 3069.1 \\
lee703q5q & visit 3 & 2021-03-18 & 08:12:51 & 3069.0 \\
lee703q7q & visit 3 & 2021-03-18 & 09:47:35 & 3069.1 \\
lee704t0q & visit 4 & 2021-03-19 & 04:50:17 & 2416.1 \\
lee704t7q & visit 4 & 2021-03-19 & 06:25:28 & 3069.2 \\
lee704t9q & visit 4 & 2021-03-19 & 08:01:35 & 3069.1 \\
lee704tbq & visit 4 & 2021-03-19 & 09:35:54 & 3069.1 \\
lee705xkq & visit 5 & 2021-03-20 & 04:38:32 & 2416.1 \\
lee705xqq & visit 5 & 2021-03-20 & 06:13:43 & 3069.1 \\
lee705xtq & visit 5 & 2021-03-20 & 07:48:56 & 3069.1 \\
\hline
\end{tabular}
%\caption{\label{table:fitsheader_COS} Observation log for the \hst -COS data. Programme ID: 16182, PI: L. Oskinova}
\end{table}

%%%%%%%%%%%%%%%%%%%%%%%%%%%%%%%%%%%%%%%%%%%%%%%%%%%%%%%%%
%%%%% TABLA DATOS FOCAS
%%%%%%%%%%%%%%%%%%%%%%%%%%%%%%%%%%%%%%%%%%%%%%%%%%%%%%%%%
\begin{table}[h]
%\caption{\label{table:fitsheader_SUBARU}SUBARU-FOCAS observation.log}
%\setlength{\tabcolsep}{0.5\tabcolsep}
\caption{\label{table:fitsheader_SUBARU} Observation log for the \textit{Subaru}-FOCAS data. Proposal ID: o11104, PI: Y. Ueda}
\centering
\begin{tabular}{lccccc}
\hline\hline
Name & Observation date & Observation time & Exposure time & Airmass & Seeing \\
 & UT (yyyy+mm-dd) & UT (hh:mm:ss) & (s) & & \\
 \hline
FCSE00122828 & 2011-02-26 &     06:46:57.25 &    1200 &  1.65 &   0.42 \\
FCSE00122830 & 2011-02-26 &    07:07:21.89 &    1200 &  1.62 &   0.44 \\
FCSE00122836 & 2011-02-26 &    07:30:28.82 &    1200 &  1.60 &   0.47 \\
FCSE00122838 & 2011-02-26 &    07:50:53.25 &    1200 &  1.59 &   0.49 \\
FCSE00123230 & 2011-02-28 &    06:26:55.48 &    1200 &  1.67 &   0.34 \\
FCSE00123232 & 2011-02-28 &    06:47:20.22 &    1200 &  1.64 &   0.57 \\
FCSE00123238 & 2011-02-28 &    07:10:18.19 &    1200 &  1.61 &   0.40 \\
FCSE00123240 & 2011-02-28 &    07:30:42.50 &    1200 &  1.59 &   0.55 \\
FCSE00123246 & 2011-02-28 &    07:53:49.08 &    1200 &  1.58 &   0.45 \\
FCSE00123648 & 2011-03-01 &    06:06:37.84 &    1200 &  1.70 &   0.32 \\
FCSE00123650 & 2011-03-01 &    06:27:02.54 &    1200 &  1.66 &   0.28 \\
FCSE00123656 & 2011-03-01 &    06:49:58.92 &    1200 &  1.63 &   0.32 \\
FCSE00123658 & 2011-03-01 &    07:10:23.37 &    1200 &  1.61 &   0.29 \\
FCSE00123664 & 2011-03-01 &    07:33:14.48 &    1200 &  1.59 &   0.28 \\
\hline
\end{tabular}
%\caption{\label{table:fitsheader_SUBARU} Observation log for the \textit{Subaru}-FOCAS data. Proposal ID: o11104, PI: Y. Ueda}
\end{table}

\clearpage
\newpage

%%%%%%%%%%%%%%%%%%%%%%%%%%%%%%%%%%%%%%%%%%%%%%%%%%%%%%%%%
%%%%% TABLA DATOS OSIRIS
%%%%%%%%%%%%%%%%%%%%%%%%%%%%%%%%%%%%%%%%%%%%%%%%%%%%%%%%%
\begin{table*}[h!]
\caption{\label{table:fitsheader_OSIRIS}Observation log for the GTC-OSIRIS data. Programme IDs: GTC20-11B and GTC38-10B, PI: F. Vilardell}
\centering
\begin{tabular}{lcccccc}
\hline\hline
 Name &  Programme ID &  O. Block &  Observation date & Observation time &   Exposure time &  Airmass\\
  &  &  &  UT (yyyy-mm-dd) & UT (hh:mm:ss) &  (s) & \\
\hline
80487	&	 GTC38-10B	&	1	&	 2010-11-11 & 03:52:31.2	&	2533	&	 1.39	\\
80488	&	&	&	 2010-11-11 & 04:35:27.0	&	2533	&	 1.36	\\
80489	&	&	&	 2010-11-11 & 05:18:22.4	&	2533	&	 1.34	\\
80490	&	&	&	 2010-11-11 & 06:01:17.6	&   2533	&	 1.35	\\
88587	&	&	2	&	 2011-01-02 & 03:24:22.7	&	2533	&	 1.37	\\
88588	&	&	&	 2011-01-02 & 04:07:18.2	&	2533	&	 1.42	\\
88589	&	&	&	 2011-01-02 & 04:50:13.4	&	2533	&	 1.49	\\
88590	&	&	&	 2011-01-02 & 05:33:08.9	&	2533	&	 1.58	\\
88695	&	&	3	&	 2011-01-02 & 22:38:30.7	&	2533	&	 1.56	\\
88696	&	&	&	 2011-01-02 & 23:21:26.2	&	2533	&	 1.47	\\
88697	&	&	&	 2011-01-03 & 00:04:21.3	&	2533	&	 1.41	\\
88698	&	&	&	 2011-01-03 & 00:47:16.8	&	2533	&	 1.37	\\
89022	&	&	4	&	 2011-01-06 & 02:22:19.5	&	2533	&	 1.35	\\
89023	&	&	&	 2011-01-06 & 03:05:15.1	&	2533	&	 1.37	\\
89024	&	&	&	 2011-01-06 & 03:48:10.5	&	2533	&	 1.41	\\
89025	&	&	&	 2011-01-06 & 04:31:06.1	&	2533	&	 1.48	\\
89496	&	&	5	&	 2011-01-09 & 23:00:10.5	&	2533	&	 1.46	\\
89497	&	&	&	 2011-01-09 & 23:43:05.7	&	2533	&	 1.40	\\
89498	&	&	&	 2011-01-10 & 00:26:01.0	&	2533	&	 1.36	\\
89499	&	&	&	 2011-01-10 & 01:08:56.3	&	2533	&	 1.34	\\
153837	&	 GTC20-11B	&	1	&	 2011-11-30 & 01:55:54.8	&	2533	&	 1.52	\\
153838	&	&	&	 2011-11-30 & 02:38:50.1	&	2533	&	 1.44	\\
153839	&	&	&	 2011-11-30 & 03:21:45.6	&	2533	&	 1.39	\\
153840	&	&	&	 2011-11-30 & 04:04:41.0	&	2533	&	 1.35	\\
174337	&	&	2	&	 2012-01-23 & 00:44:33.6	&	2533	&	 1.35	\\
174338	&	&	&	 2012-01-23 & 01:27:29.1	&	2533	&	 1.34	\\
174339	&	&	&	 2012-01-23 & 02:10:24.7	&	2533	&	 1.35	\\
174340	&	&	&	 2012-01-23 & 02:53:20.1	&	2533	&	 1.38	\\
175417	&	&	3	&	 2012-01-26 & 22:00:05.2	&	2533	&	 1.53	\\
175418	&	&	&	 2012-01-26 & 22:43:00.6	&	2533	&	 1.45	\\
175419	&	&	&	 2012-01-26 & 23:25:56.0	&	2533	&	 1.40	\\
175420	&	&	&	 2012-01-27 & 00:08:51.6	&	2533	&	 1.36	\\
175423	&	&	&	 2012-01-27 & 01:02:02.6	&	6	&	 1.34	\\
175592	&	&	4	&	 2012-01-28 & 00:01:59.4	&	2533	&	 1.36	\\
175593	&	&	&	 2012-01-28 & 00:47:08.0	&	2533	&	 1.34	\\
175597	&	&	&	 2012-01-28 & 02:07:29.0	&	2533	&	 1.36	\\
175598	&	&	&	 2012-01-28 & 02:51:04.0	&	2533	&	 1.40	\\
175730	&	&	 0005A1	&	 2012-01-29 & 21:13:33.9	&	2533	&	 1.62	\\
175731	&	&	&	 2012-01-29 & 21:59:04.6	&	2533	&	 1.51	\\
175732	&	&	&	 2012-01-29 & 22:42:54.6	&	2533	&	 1.44	\\
175733	&	&	&	 2012-01-29 & 23:26:15.5	&	2533	&	 1.38	\\
\hline
\end{tabular}
\end{table*}

\end{appendix}

\end{document}